\documentclass[12pt,a4paper,dvips]{article}
\usepackage{a4p}
\usepackage{cite,mcite}
\usepackage{graphicx}
\usepackage{physics }
\usepackage{l3_title,ifthen,Lep}
\journalname{Phys. Lett. B}
\date{November 13, 2000}
\preprint{2000-143}
\Lep{2}
\newlength{\capindent}
\newlength{\capwidth}
\newlength{\figwidth}
\newcommand{\icaption}[2][!*!,!]{\hspace*{\capindent}%
  \begin{minipage}{\capwidth}
    \ifthenelse{\equal{#1}{!*!,!}}%
      {\caption{#2}}%
      {\caption[#1]{#2}}
  \end{minipage}}
\begin{document}
\begin{titlepage}
\title{Search for Excited Leptons in \boldmath{$\rm e^+ e^-$} Interactions 
       at \boldmath{$\sqrt{s}=192 - 202 \GeV$} }

\author{The L3 Collaboration}

%
%
\begin{abstract}
Excited leptons are searched for using the L3 detector at LEP.
The data collected at centre-of-mass energies in the range from 192 up 
to $202 \GeV$ correspond to a total luminosity of 233 $\rm pb^{-1}$.
No evidence of either pair production of excited leptons, nor of single 
production is found.  From the searches for pair produced excited leptons, 
lower mass limits close to the kinematic limit are set. 
From the searches for singly produced excited leptons, upper limits on 
their couplings are derived in the mass range up to $200 \GeV$. 
\end{abstract}

\submitted
\end{titlepage}

%
%
\section{Introduction} 

Several fundamental aspects of the Standard Model\cite{SM},
such as the number of families and the fermion masses 
could be explained naturally in composite models where 
leptons and quarks have substructure\cite{yellow,hagi,neus}.
The existence of excited leptons is a typical consequence of these models.

The excited leptons 
$\rm e^*,~\mu^*, ~\tau^*, ~\nu_{e}^*, ~\nu_{\mu}^*$ and $\nu_{\tau}^*$ 
were extensively searched for at the LEP\cite{L3_189,LEP} \ee collider.
First generation excited leptons were also searched for at
the HERA\cite{HERA} $\rm ep$ collider.
The increase of the LEP centre-of-mass energy, $\sqrt{s}$, up to 
$202 \GeV$ in 1999 extends the range of potential discovery of new
heavy particles.

The interactions of excited leptons are studied within a model\cite{hagi} that 
assumes spin 1/2, and isospin doublets with left and right handed components: 
$$
  {\rm L}^* =
  \left(\begin{array}{c} \nu^* \\ \ell^* \end{array}\right)_L + 
  \left(\begin{array}{c} \nu^* \\ \ell^* \end{array}\right)_R,
$$
with $\rm \ell = e, \mu, \tau$ and $\rm \nu = \nu_{e}, \nu_{\mu}, \nu_{\tau}$.
Excited leptons are expected to decay into their ground states by 
radiating a photon ($\rm \ell^* \longrightarrow \ell \gamma$, 
$\rm \nu^* \longrightarrow \nu \gamma$),
via charged-current decay ($\rm \ell^* \longrightarrow \nu W$,
$\rm \nu^* \longrightarrow \ell W$) 
or via weak neutral-current decay ($\rm \ell^* \longrightarrow \ell Z$, 
$\rm \nu^* \longrightarrow \nu Z$).

At \ee~colliders, excited leptons can be produced either in pairs 
(\ee $\rm \longrightarrow \ell^*\ell^*$,
\ee $\rm \longrightarrow \nu^*\nu^*$) or singly 
(\ee $\rm \longrightarrow \ell \ell^*$, \ee $\rm \longrightarrow \nu \nu^*$).
In pair production, the coupling of the excited leptons to the gauge bosons
is described by the effective Lagrangian:
$${\cal L}_{{\rm L}^*{\rm L}^*} = \bar{\rm L}^* \gamma^{\mu} \left( g  
{\vec\tau \over 2} \vec W_{\mu} + {g'} Y B_{\mu} \right) {\rm L}^*. $$
Here $\gamma^{\mu}$ are the Dirac matrices, $g$ and $g'$ are the Standard 
Model SU(2) and U(1) coupling constants, $\vec\tau$ denotes the Pauli 
matrices, $Y=-1/2$ is the hypercharge, and $\vec W$ and $B$ are the gauge 
fields associated with the SU(2) and U(1) groups respectively.
The couplings of excited leptons to the gauge bosons are fixed by their
quantum numbers, and thus the cross section for pair production 
depends only on the mass of the excited lepton and the centre-of-mass energy. 

The single production of excited leptons as well as their decay are
described by means of an effective Lagrangian of the form:
$$
{\cal L}_{{\rm L}^*{\rm L}} = 
{1 \over \Lambda} {\bar {\rm L}^*} \sigma^{\mu\nu}  
\left( g  f  \frac{\vec\tau}{2} \partial_{\mu}{\vec W_{\nu}} + 
       g' f' Y \partial_{\mu}B_{\nu} \right)  
       \frac{1-\gamma^5}{2} {\rm L} + {\rm h.c.},$$
where 
$\sigma^{\mu\nu} = \frac{\textstyle i}{\textstyle 2}
[\gamma^{\mu},\gamma^{\nu}]$,
$\Lambda$ is the scale of new physics and $f$ and $f'$ are the couplings
associated with SU(2) and U(1) respectively. 
The production rate of single excited leptons depends on their mass, 
$\sqrt{s}$, $f/\Lambda$ and $f'/\Lambda$. 
For a given mass and centre-of-mass energy, the polar angle distribution 
is determined by the relative value of the couplings $f$ and $f'$. 
The decay fractions of excited leptons into standard leptons plus 
gauge bosons are also determined by the relative value of 
$f$ and $f'$\cite{neus}. Table~\ref{tab:bratio} shows the branching ratios 
for two choices of $f$ and $f'$ and for different excited lepton masses.

Pair production searches are sensitive to excited leptons
of mass up to values close to the kinematic limit, 
{\it i.e.} the beam energy, $E_{beam}$.
Searches for single production extend the sensitivity to the mass range above 
the beam energy up to the centre-of-mass energy.

In the case of single production of excited electrons, the polar angle 
distribution depends critically on the relative values of $f$ and $f'$. 
If $f=f'$, the $t$-channel photon exchange gives a large contribution at 
low polar angle with the electron escaping through the beam pipe.
If $f=-f'$, the $\rm e^* e \gamma$ vertex is not allowed, 
and only the $\rm Z$ mediated $s$- and $t$-channels contribute, 
giving rise to events with an electron visible in the detector. Thus, the
efficiency for $\rm e^*$ selection is calculated for both scenarios: $f=f'$
and $f=-f'$. For all other flavours of excited leptons, the detection
efficiency depends only slightly on the coupling assumption.

%
%
\section{Data Sample and Event Simulation}

The data sample analysed was collected in 1999 with the L3 
detector\cite{l3-detector} at four different centre-of-mass energies,
$191.6 \GeV$, $195.5 \GeV$, $199.5 \GeV$ and $201.7 \GeV$, with corresponding
integrated luminosities of 29.7 $\rm pb^{-1}$, 83.7 $\rm pb^{-1}$,
82.8 $\rm pb^{-1}$ and 37.0 $\rm pb^{-1}$. 

Standard Model processes are simulated by different Monte Carlo programs.
Radiative Bhabha events are generated using BHWIDE\cite{bhwide} and 
TEEGG\cite{tee}. 
For other radiative dilepton events, $\mu\mu\gamma$, $\tau\tau\gamma$ 
and $\nu\nu\gamma$, the KORALZ\cite{koralz} generator is used. 
KK2F\cite{kk2f} is used for  $\rm q\bar{q}(\gamma)$ production. 
The GGG\cite{ggg} Monte Carlo is used for final states with only photons. 
Four-fermion processes are generated with EXCALIBUR\cite{exca} for 
$\rm q \bar{q} e \nu$, $\ell\ell\ell\ell$ and $\ell\ell\nu\nu$ final states. 
PYTHIA\cite{pythia} is used for final states coming from Z pair and
Zee production not covered with EXCALIBUR.
Further final states from W pair production are modelled with 
KORALW\cite{koralw}.
The production of hadrons in two-photon interactions is described
by PHOJET\cite{phojet} while DIAG36\cite{diag36} is used to model lepton
production.

Several samples for different flavours and masses of excited leptons are 
generated in order to optimise the selections and estimate the efficiencies. 
Pair produced excited leptons are generated with a mass of $95 \GeV$ 
which is close to the expected limits. In the case of single production,
excited leptons with masses 100, 150 and $195 \GeV$ are generated, and 
a linear interpolation is used to estimate the selection efficiency for 
masses in the range from 90 to $200 \GeV$.
The expected differential cross sections\cite{hagi} and decay 
modes\cite{neus} are modelled.  Initial state radiation is not implemented 
in the generation, but is taken into account in cross section calculations.

The L3 detector response is simulated for all Monte Carlo samples
using the GEANT program\cite{geant}, which takes into account the 
effects of energy loss, multiple scattering and showering in the 
detector. Time dependent inefficiencies, as monitored during the 
data taking period, are reproduced in these simulations.

%
%
\section{Selection Strategy}

The selection of candidate events for excited lepton production is performed
with criteria similar to those used at $\sqrt{s} = 189 \GeV$\cite{L3_189},
for the most significant experimental signatures.
These criteria are modified to follow the evolution of the signal topology,
that manifests a larger boost of the decay particles with the higher
$\sqrt{s}$.
In addition, some new selections are performed to complement the search 
in decay channels which were not investigated previously.
All selections are based on the identification of the decay products 
of the excited leptons. 

Photons and electrons are identified as clusters with energy above $1 \GeV$
in the BGO electromagnetic calorimeter.  Photons must be isolated 
whereas electrons must be associated to a charged track.
In order to be well inside the angular acceptance of the central 
tracking chamber, the polar angle of electrons and photons must
satisfy $|\cos{\theta}| < 0.95$.

Muons are reconstructed either from tracks in the muon spectrometer
pointing to the interaction vertex and in time with the event,
or via energy depositions in the calorimeters consistent with a minimum 
ionising particle.

Tau leptons are reconstructed from low multiplicity hadronic jets,
or from identified electrons or muons. 
The neutrino from the tau decay carries away part
of the energy, giving rise to events with missing energy along the
tau direction. In some selections where a kinematic fit is applied,
this signature is imposed by requiring the fitted energy of the tau 
to be greater than the visible energy of the corresponding jet or lepton.

To reduce the background from the radiative return to the Z and 
from two-photon interactions, the missing 
momentum is required to point away from the beam axis
in most of the selections.

The discovery potential is maximised for the topologies in which it 
is possible to reconstruct the invariant mass of the excited leptons.
Pair production and single production of excited leptons are discussed below.

%
%
\section{Pair Production}

The search for pair produced excited leptons is optimised for excited 
leptons of masses close to the kinematic limit.
For excited leptons of mass up to $100 \GeV$, the neutral-current
decay branching ratio is always below 15\%, and depends
on the mass and the relative value of $f$ and $f'$. Thus, only
the radiative and the charged-current decays are considered.
The production cross section can reach a few pb, depending on the 
excited lepton mass. 
For example at $\sqrt{s} = 200 \GeV$ and an excited lepton mass of $95 \GeV$, 
the cross section for pair production is 1.1 pb for $\ell^*$ and
0.7 pb for $\nu^*$.

Table~\ref{tab:selections} lists the selections and
summarises the efficiencies and the number of events found in data
and expected from Standard Model background processes.
The relative importance of the selections is given by the coupling assumption.

%
%
\subsection{Radiative Decays}


The radiative decays $\rm \ell^* \longrightarrow \ell \gamma$ and
$\rm \nu^* \longrightarrow \nu \gamma$ provide the maximum sensitivity
in the $f=f'$ and the $f=-f'$ scenarios respectively.
Event selection for the 
\ee$\rm\longrightarrow \ell^*\ell^* \longrightarrow \ell\ell\gamma\gamma$
process requires low multiplicity events with two photons and two same
flavour charged leptons. 
To reduce the background from radiative dilepton production, at least
one of the photons is required to lie in the central region of the 
detector ($|\cos{\theta_\gamma}| < 0.75$, where $\theta_\gamma$ is the 
photon polar angle).
After a kinematic fit, which imposes energy and momentum conservation, 
the difference of the lepton-photon invariant masses is required to be 
smaller than $10 \GeV$ and their sum greater than $100 \GeV$.


The selection of the
\ee$\rm\longrightarrow \nu^*\nu^* \longrightarrow \nu\nu\gamma\gamma$
channel requires two photons in the central region of the detector with 
energies ${\textstyle E_{\gamma}}$ in the range
$0.2 < {\textstyle E_{\gamma}}/{\textstyle E_{beam}} < 0.8$.
There should be no tracks in the central tracker nor in the muon spectrometer,
and the energy of any other calorimetric cluster should be less than $5 \GeV$.
The background from the \ee$\longrightarrow \gamma\gamma(\gamma)$ 
process is efficiently reduced by requiring the photon acoplanarity 
to be greater than $10^\circ$.

%
%
\subsection{Charged-Current Decays}


The charged-current decays $\rm \ell^* \longrightarrow \nu W$ and
$\rm \nu^* \longrightarrow \ell W$ are dominant in the $f=-f'$ and the 
$f=f'$ scenarios respectively.
Final states from
\ee$\rm\longrightarrow \ell^*\ell^* \longrightarrow \nu\nu W W$ 
are experimentally similar to Standard Model W pair production. 
To achieve a high signal efficiency, pair production of excited leptons 
is searched for as an enhancement of the W pair-like event rate.
A combination of four selections, named
$\rm qqqq$, $\rm qqe\nu$, $\rm qq\mu\nu$ and $\rm qq\tau\nu$, 
selects such events.

\begin{figure}[htb]
  \begin{center}
    \includegraphics[width=0.49\textwidth]{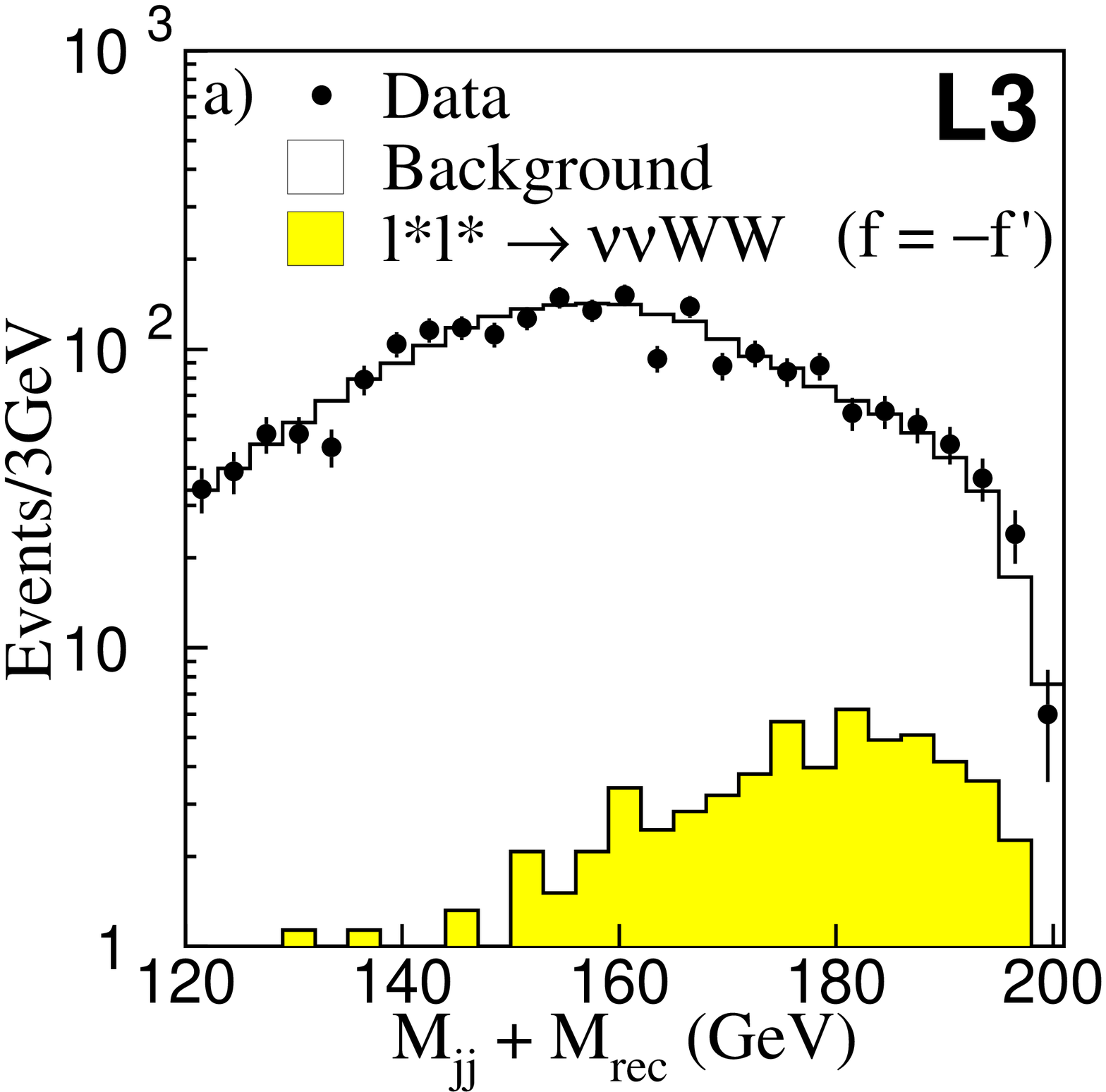}
    \includegraphics[width=0.49\textwidth]{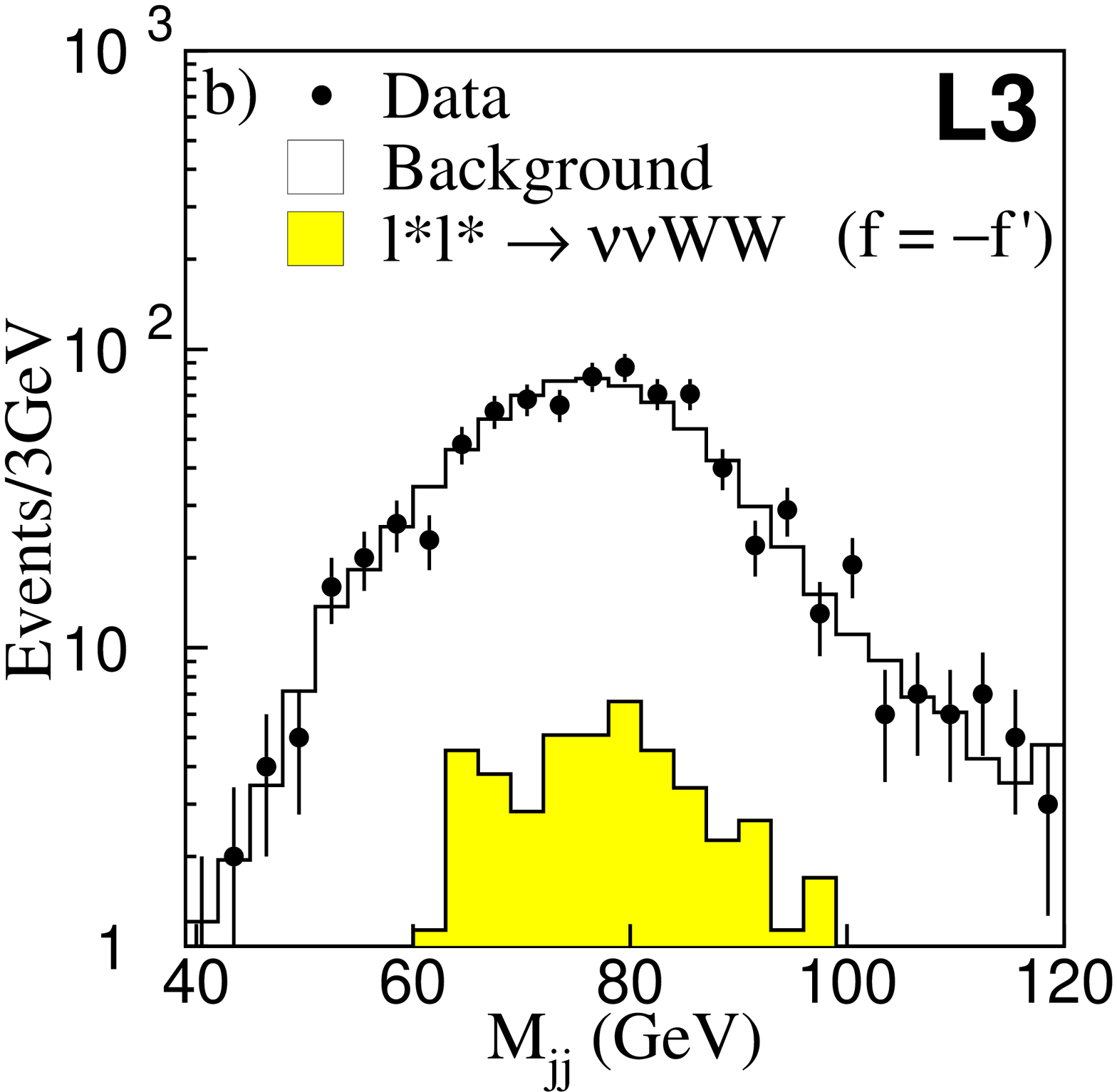}
    \includegraphics[width=0.49\textwidth]{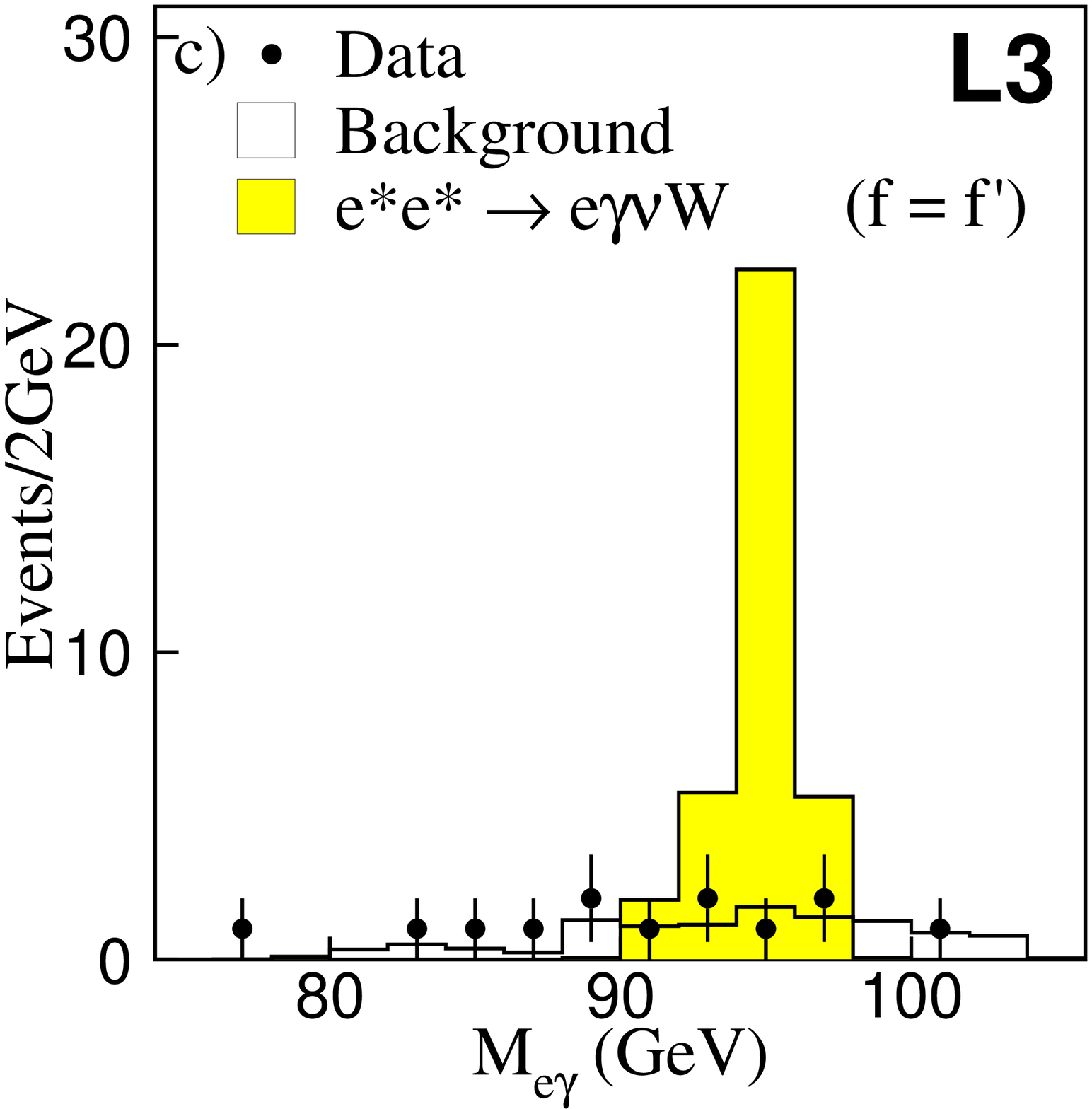}
    \includegraphics[width=0.49\textwidth]{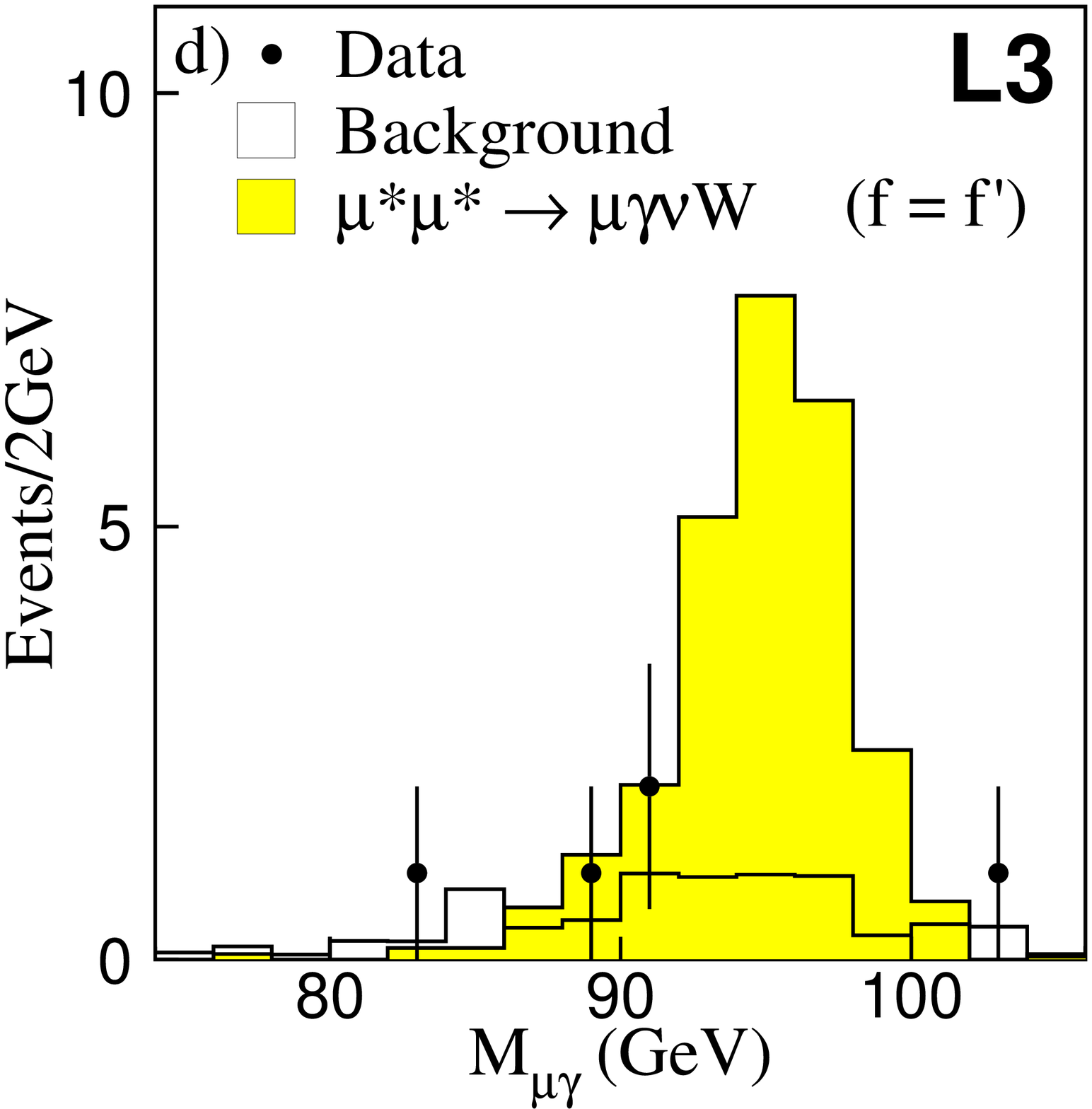}
  \end{center}
  \icaption{Distributions of a) the sum of invariant, $M_{jj}$, and recoil
            mass, $M_{rec}$, for two jets in the $\rm qqqq$ selection;
            b) the hadronic invariant mass, $M_{jj}$, in the 
            $\rm qq\ell\nu$ selection;
            c) the electron photon invariant mass in the 
            $\rm e\gamma\nu W$ selection and 
            d) the muon photon invariant mass in the
            $\rm \mu\gamma\nu W$ selection.
            The expected signal for an excited lepton with a 
            mass of $95 \GeV$ is shown together with data and 
            Standard Model background; its luminosity averaged cross 
            section is 0.93 pb.
  \label{fig:pair}}
\end{figure}

For the $\rm qqqq$ selection, events with at least three hadronic jets 
and visible energy greater than $140 \GeV$ are selected.
For at least one pair of jets, their invariant mass must be in the 
range between $50 \GeV$ and $110 \GeV$, the recoiling mass against these 
two jets must be greater than $50 \GeV$ and the sum of invariant plus 
recoil masses must be greater than $120 \GeV$.
Figure~\ref{fig:pair}a shows the sum of invariant and recoil masses
for this pair of jets. 
The signal distribution corresponds to the $f=-f'$ coupling assumption.


For the $\rm qq\ell\nu$ selections, high multiplicity events with 
missing momentum greater than $20 \GeV$, visible energy below 
$0.8 \sqrt{s}$ and an isolated lepton are retained. 
The invariant mass of the hadronic system must be compatible
with the W mass.
In the $\rm qq\tau\nu$ selection, the $\rm q \bar{q} (\gamma)$ background 
is further reduced by requiring the missing momentum along the 
longitudinal direction to be smaller than $55 \GeV$, and rejecting 
events with high energy photons.
Figure~\ref{fig:pair}b shows the hadronic invariant mass
for the  $\rm qq\ell\nu$ selected events.


The experimental signatures of the process
\ee$\rm\longrightarrow \nu^*\nu^* \longrightarrow \ell\ell W W$
are also similar to Standard Model W pair production. 
Thus, the $\rm qqqq$ and $\rm qq\ell\nu$ selections described above are 
used to investigate them.
Moreover, in the  $\rm e e W W$ and $\rm \mu \mu W W$ channels, the final
state leptons may be detected, providing a very clean signature. 
Therefore, for these channels the signal sensitivity is enhanced 
by requiring two additional isolated electrons or muons in the event, 
with energy in the range between $3 \GeV$ and $25 \GeV$.

%
%
\subsection{Mixed Decays}


The processes 
\ee$\rm\longrightarrow \ell^*\ell^* \longrightarrow \ell \gamma \nu W$ and
\ee$\rm\longrightarrow \nu^*\nu^*   \longrightarrow \nu  \gamma \ell W$ 
give rise to events with an energetic photon, a lepton, and the decay 
products of a W boson. Both the hadronic and the leptonic decays
of the W are considered.
The selected events must contain an energetic photon at high polar angle. 
The additional criteria for each final state are discussed below.

In the $\rm e^*e^* \longrightarrow e \gamma \nu W$ and 
$\rm \mu^*\mu^* \longrightarrow \mu \gamma \nu W$ channels, an 
energetic electron or muon is required. Moreover, the sum of 
the invariant and recoil masses of the lepton-photon pair,
{\it i.e.} the masses of the $\ell^*$ candidates, must be greater
than $160 \GeV$ and their difference smaller than $10 \GeV$ in the 
electron case and $20 \GeV$ in the muon case. 
Figures~\ref{fig:pair}c and d show the $\rm e \gamma$ and $\mu \gamma$
invariant masses for the selected events.
For the signal distributions, $f=f'$ is assumed to set the branching ratios.

For the $\rm \tau^* \tau^* \longrightarrow \tau \gamma \nu W$ and 
$\rm \nu^* \nu^* \longrightarrow \nu \gamma \ell W$ channels, 
a low multiplicity jet, an electron or a muon must be present. 
In high multiplicity events, the signature of a W boson is imposed by
requiring the invariant mass of the hadronic system to be in the
range from $60 \GeV$ to $100 \GeV$, whereas low multiplicity events must
contain only the photon and no more than two leptons.
For the $\rm \tau^* \tau^* \longrightarrow \tau \gamma \nu W$ channel, 
the invariant mass and the energy of the $\rm \nu W$ system correspond to 
the $\tau^*$ mass and the beam energy, respectively. This signature
is used by requiring the recoil mass
of the tau-photon system to exceed $80 \GeV$, and the visible 
energy of the rest of the event to be smaller than $120 \GeV$.
For the $\rm \nu^* \nu^* \longrightarrow \nu \gamma \ell W$ channel, 
after excluding the photon, the visible energy of the rest of the event 
must be smaller than $120 \GeV$ and  $70 \GeV$
for the hadronic and leptonic decays of the W respectively. 

%
%
\section{Single Production}

The search for singly produced excited leptons complements the search
for pair production in the mass range from $E_{beam}$ to $\sqrt{s}$.
The selections are optimised to cover this wide range of masses with 
high efficiency.
The radiative decays of excited leptons, as well as the charged-current 
decays and the neutral-current decays are investigated, as described below. 
Both the hadronic and the leptonic decays of W and Z bosons are considered.

Table~\ref{tab:selections} lists the selections and
summarises the efficiencies and the number of events found in data
and expected from Standard Model background processes.

%
%
\subsection{Radiative Decays}

\begin{figure}[htb]
  \begin{center}
    \includegraphics[width=0.49\textwidth]{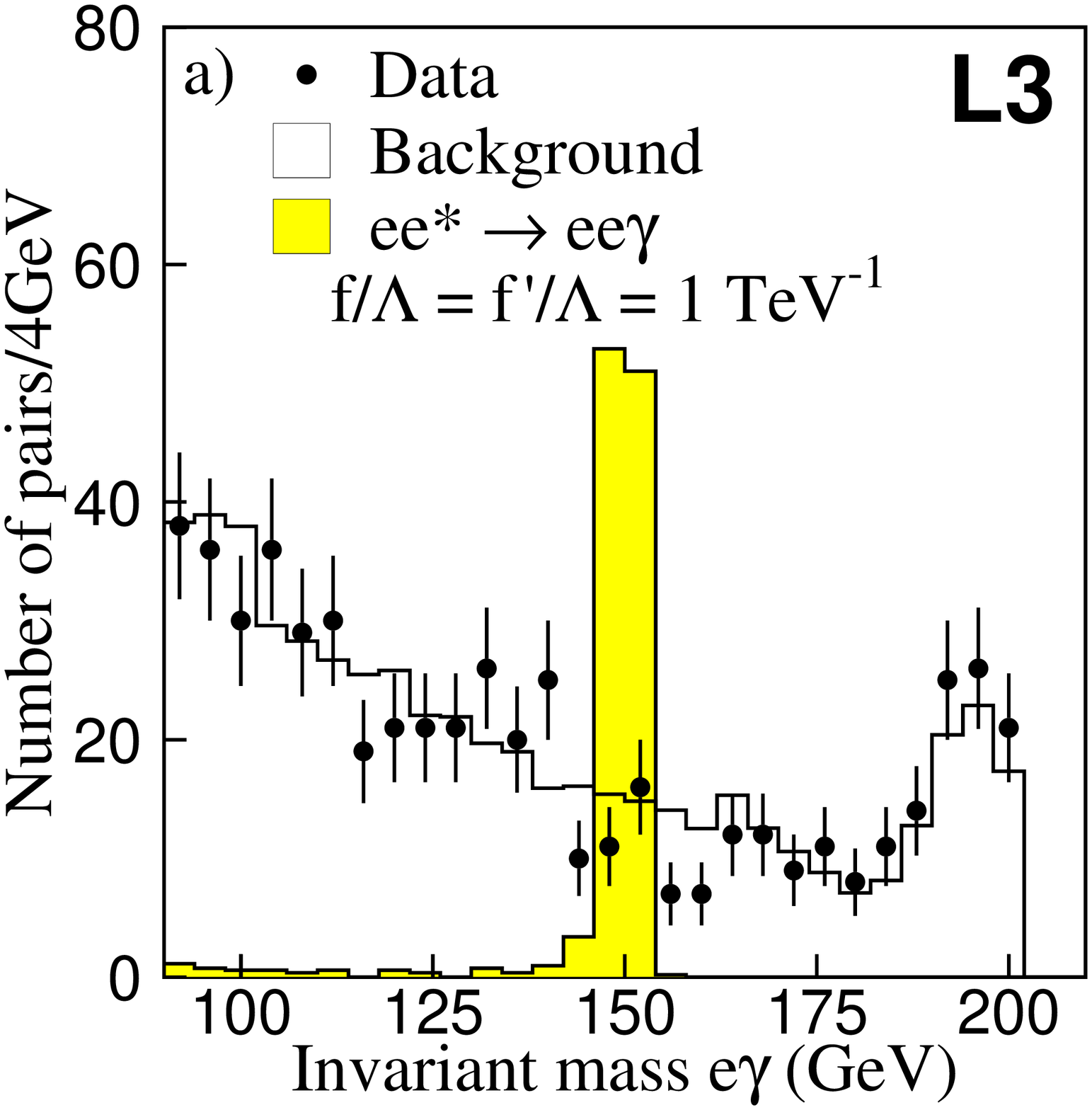}
    \includegraphics[width=0.49\textwidth]{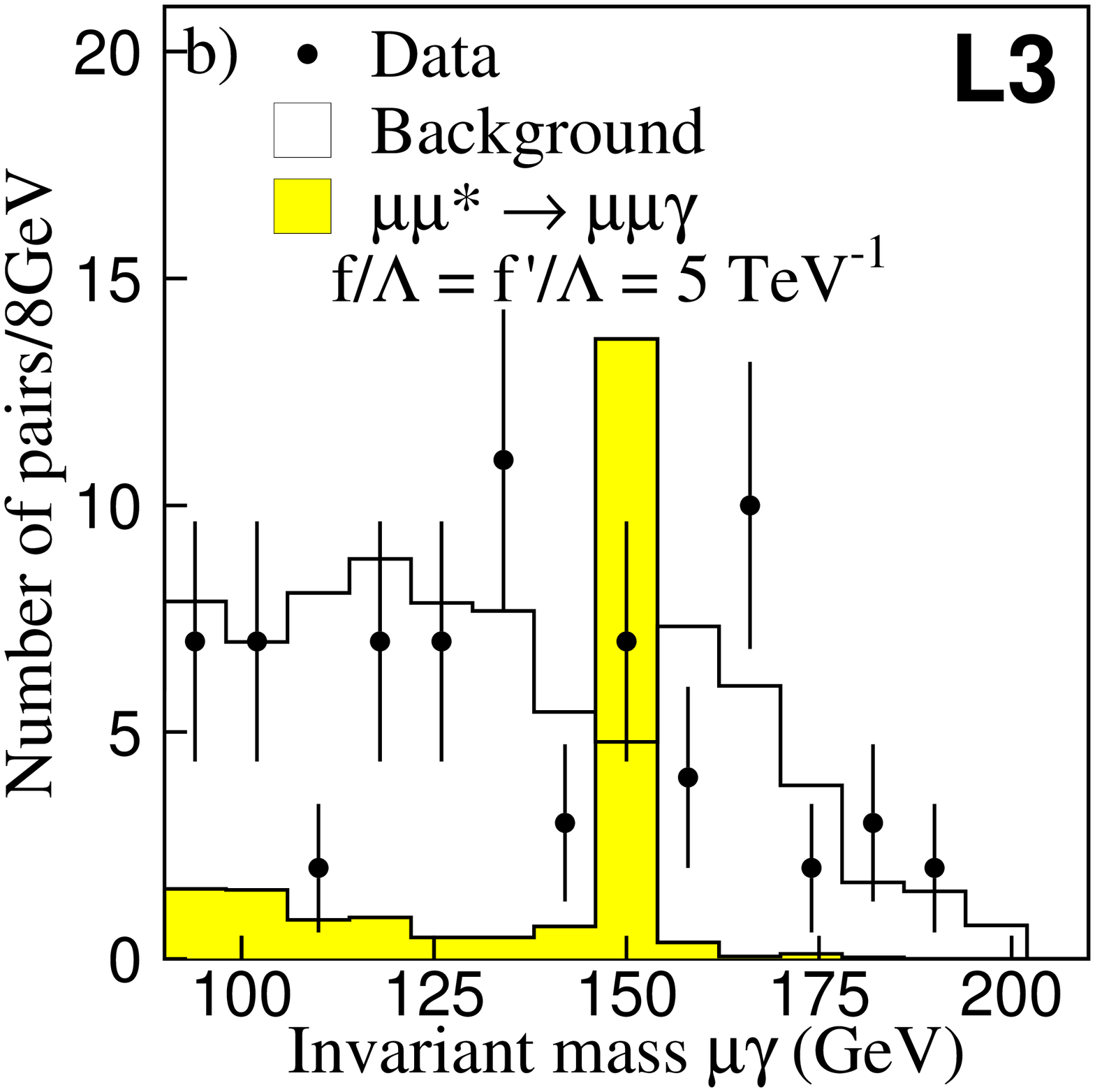}
    \includegraphics[width=0.49\textwidth]{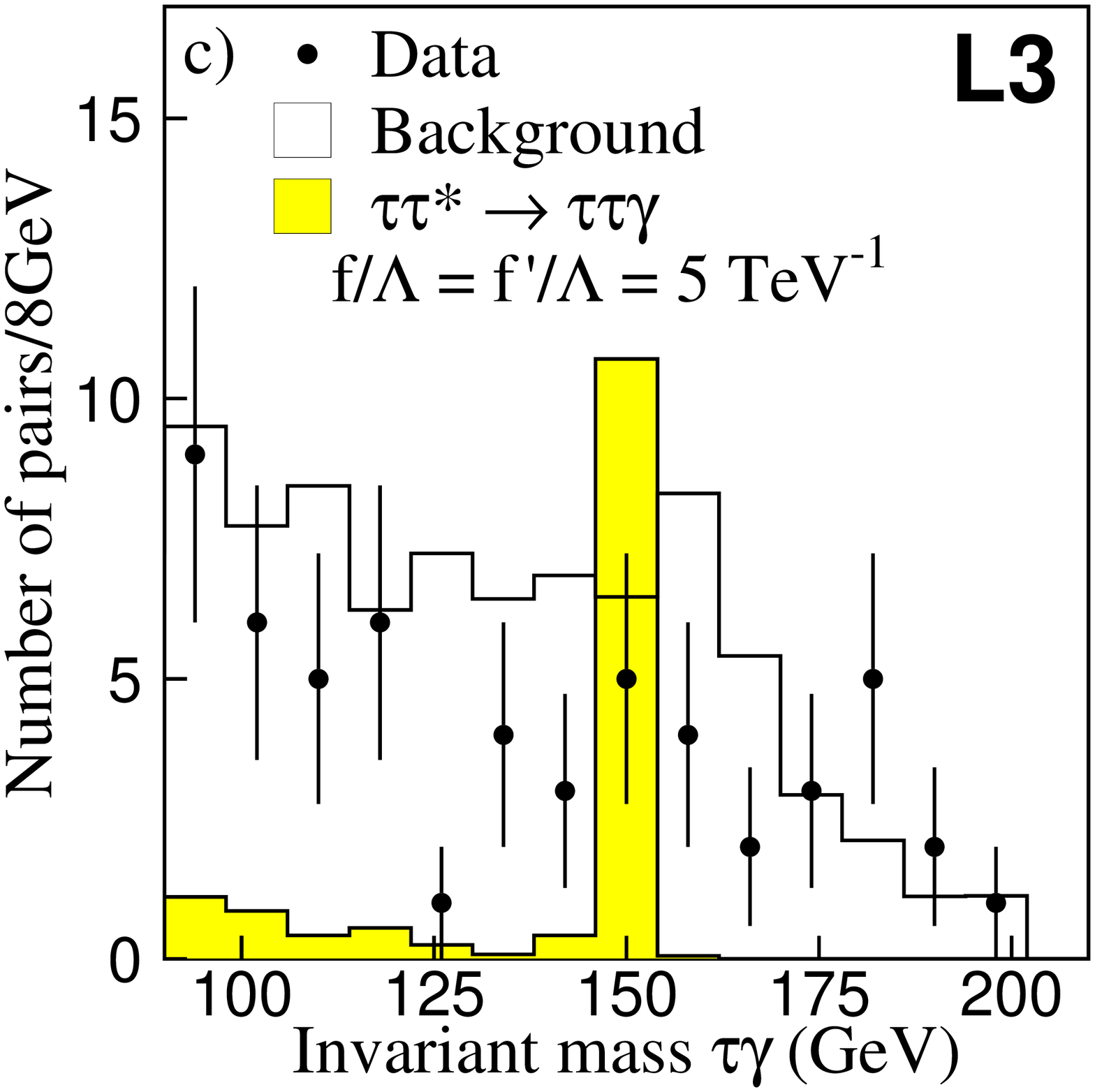}
    \includegraphics[width=0.49\textwidth]{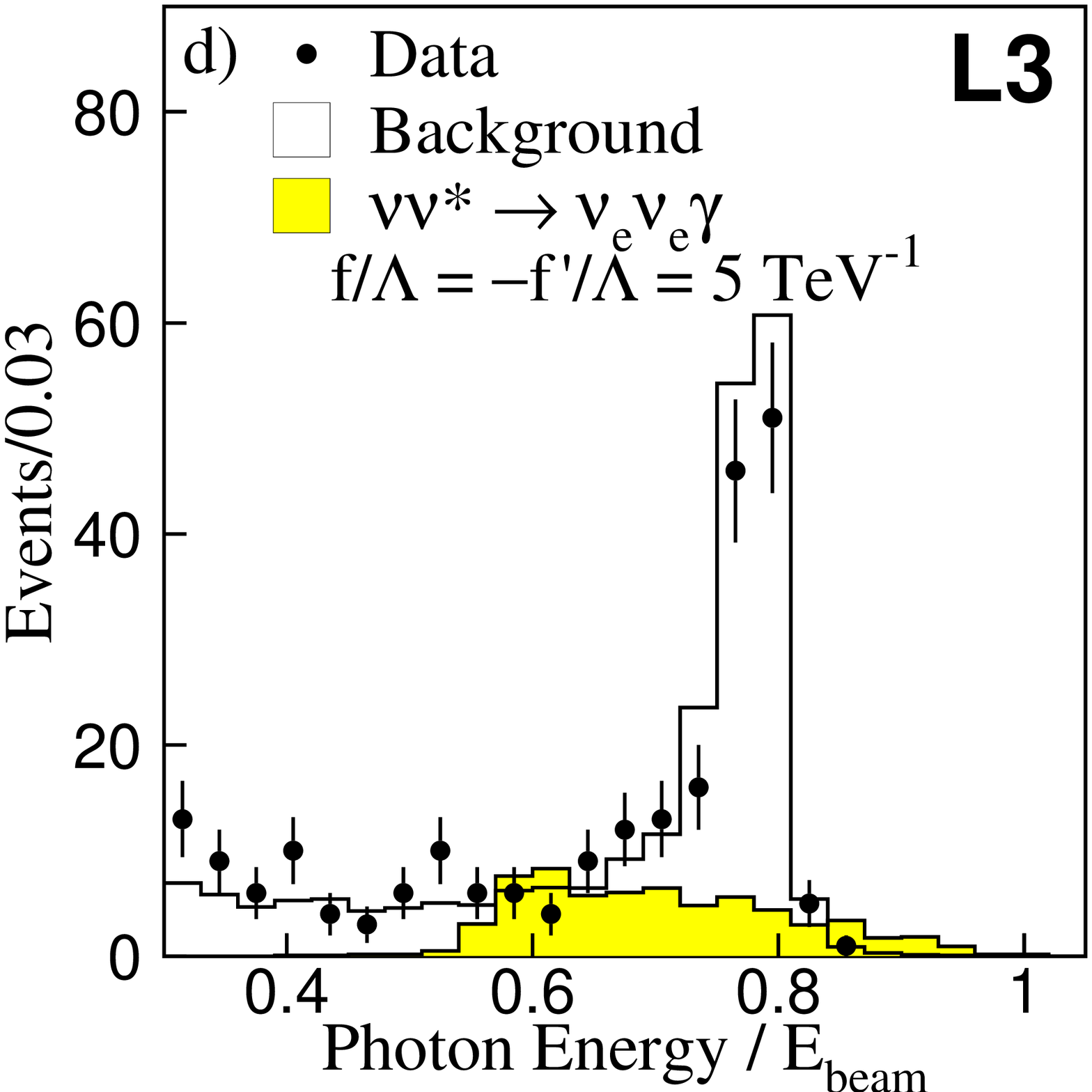}
  \end{center}
  \icaption{The invariant mass distributions for a) e$\gamma$,
           b) $\mu\gamma$, and c) $\tau\gamma$ pairs. 
           d) The normalised energy distribution
           of single photon events.
           The expected signal
           for an excited lepton with a mass of $150 \GeV$  
           is shown together with data and Standard
           Model background.
           The signals are plotted for the arbitrary choice of couplings 
           displayed in the Figures.
  \label{fig:rad}}
\end{figure}

The event selection for
\ee$\rm\longrightarrow \ell\ell^* \longrightarrow \ell\ell\gamma$ identifies
final states with two leptons and one photon with energy greater than 
$20 \GeV$.  The background from radiative dilepton production is reduced by 
requiring the photon to be in the central region of the detector,
$|\cos{\theta_\gamma}| < 0.75$. After a kinematic fit, at least one of the 
two possible $\rm \ell \gamma$ invariant masses must be greater than $70 \GeV$.
Events with just an identified electron and a photon, with invariant mass
above $70 \GeV$, are also accepted for the excited electron selection. Thus,
a high signal efficiency is kept for the signal events originating from the 
$t$-channel photon exchange where one electron escapes along the beam pipe. 
Figures~\ref{fig:rad}a-c show all combinations of lepton-photon invariant 
masses above $90 \GeV$.

Final states from 
\ee$\rm\longrightarrow \nu\nu^* \longrightarrow \nu\nu\gamma$ are selected 
by requiring  a single photon with energy greater than $0.15 \sqrt{s}$ at 
high polar angle $|\cos{\theta_\gamma}| < 0.75$.
Neither tracks in the tracking chamber, nor in the muon spectrometer should
be present.
Figure~\ref{fig:rad}d shows the distribution of the normalised photon energy.

%
%
\subsection{Charged-Current Decays}

\begin{figure}[htb]
  \begin{center}
    \includegraphics[width=0.40\textwidth]{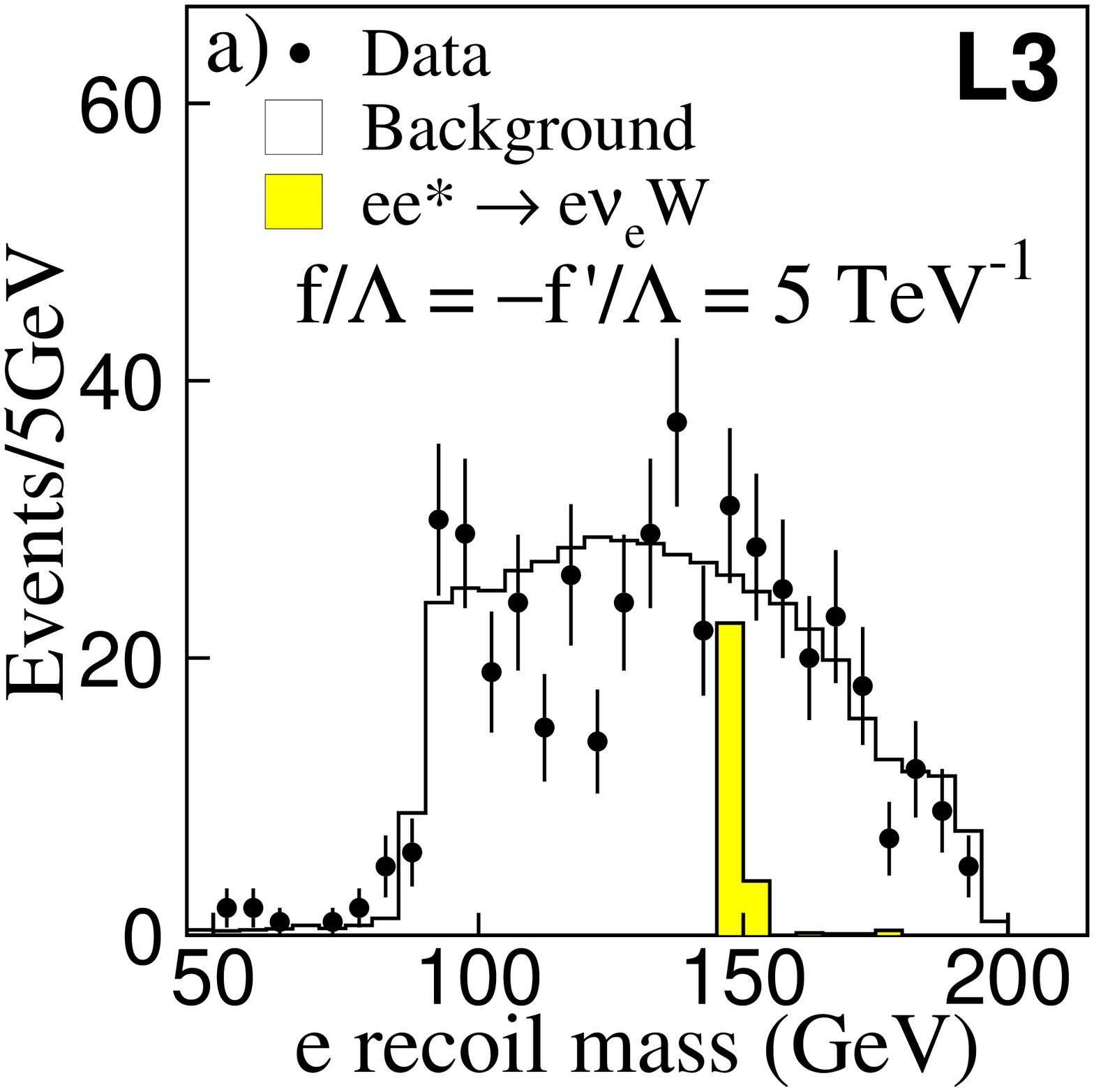}
    \includegraphics[width=0.40\textwidth]{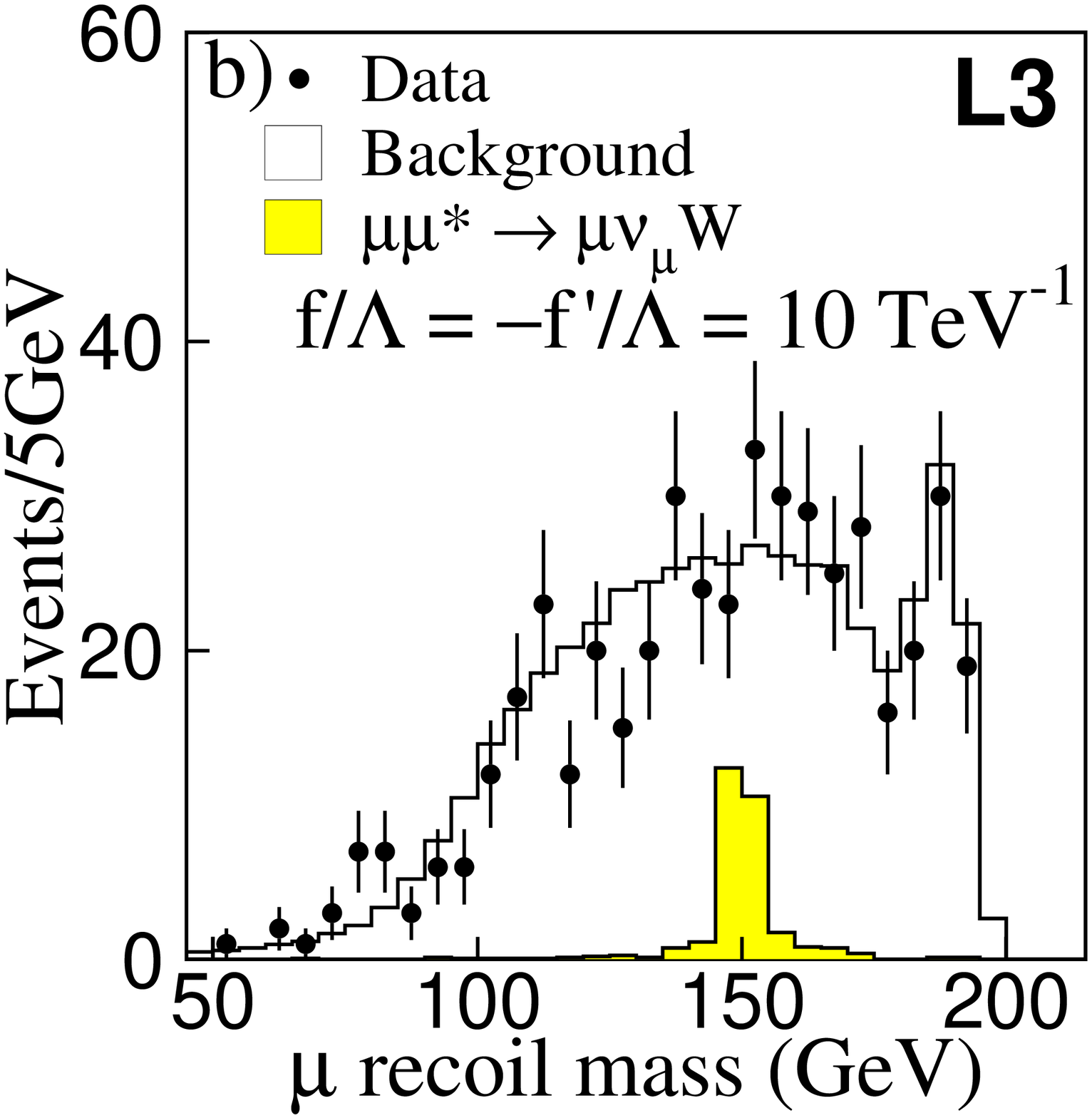}
    \includegraphics[width=0.40\textwidth]{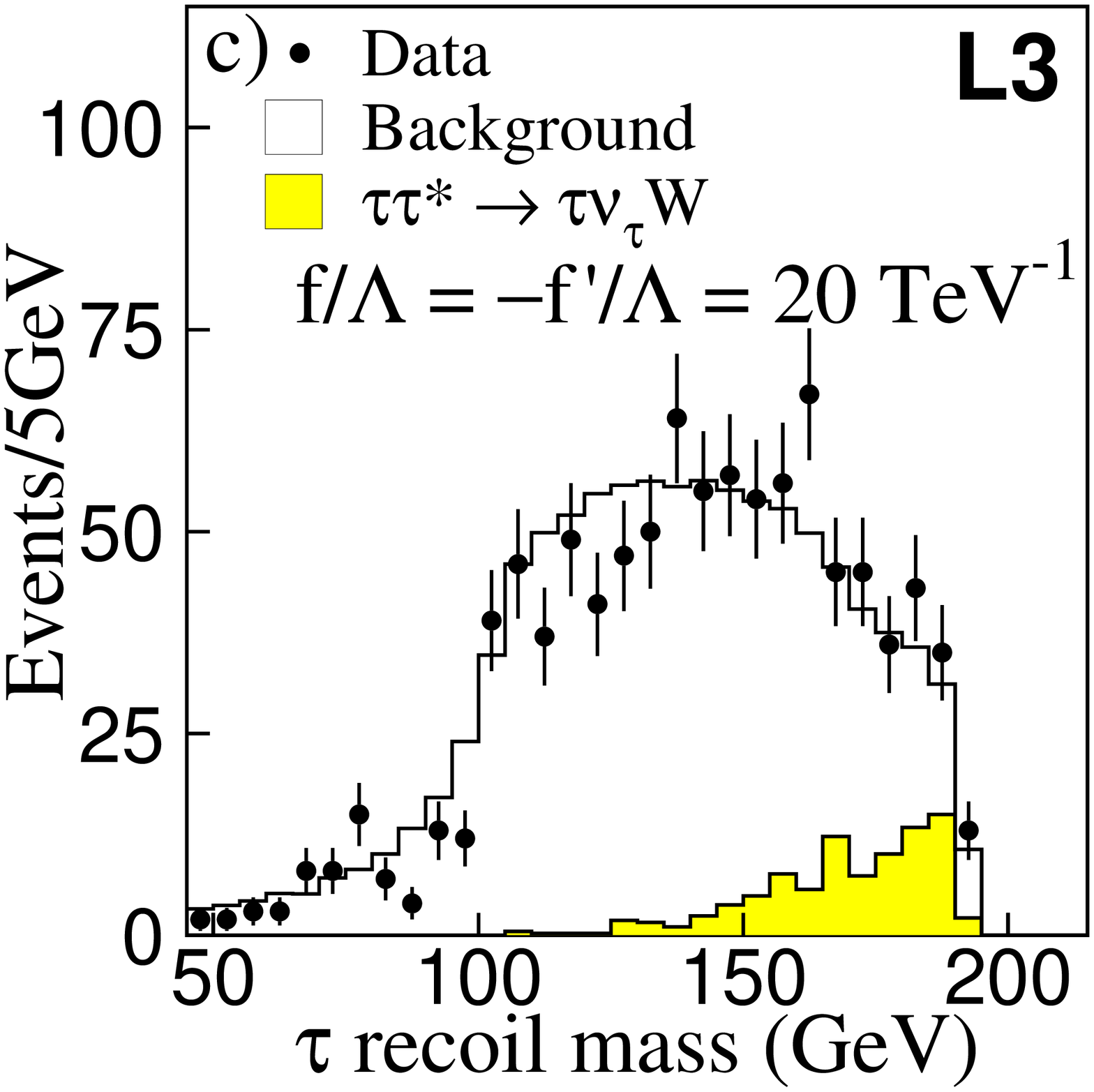}
    \includegraphics[width=0.40\textwidth]{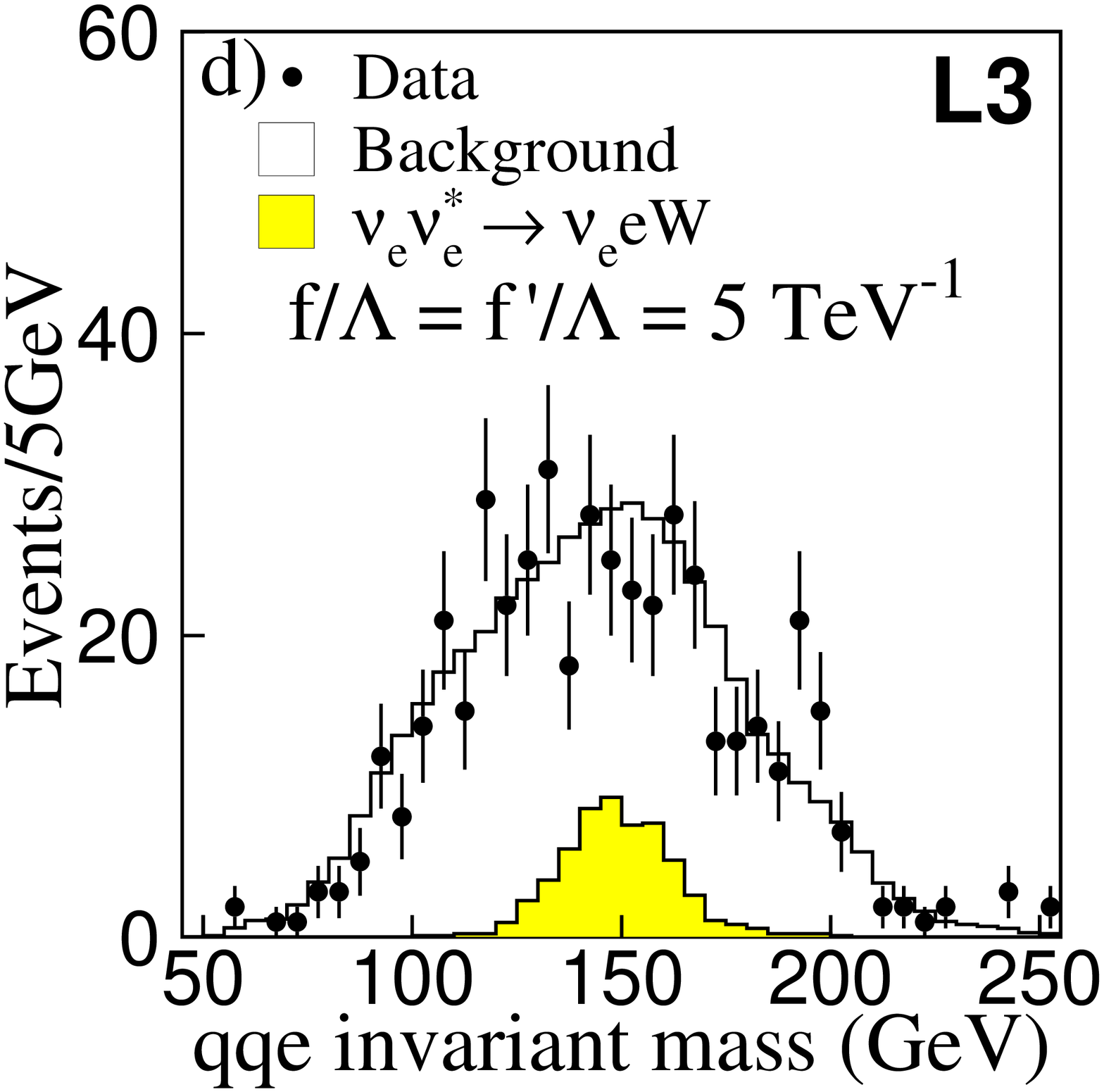}
    \includegraphics[width=0.40\textwidth]{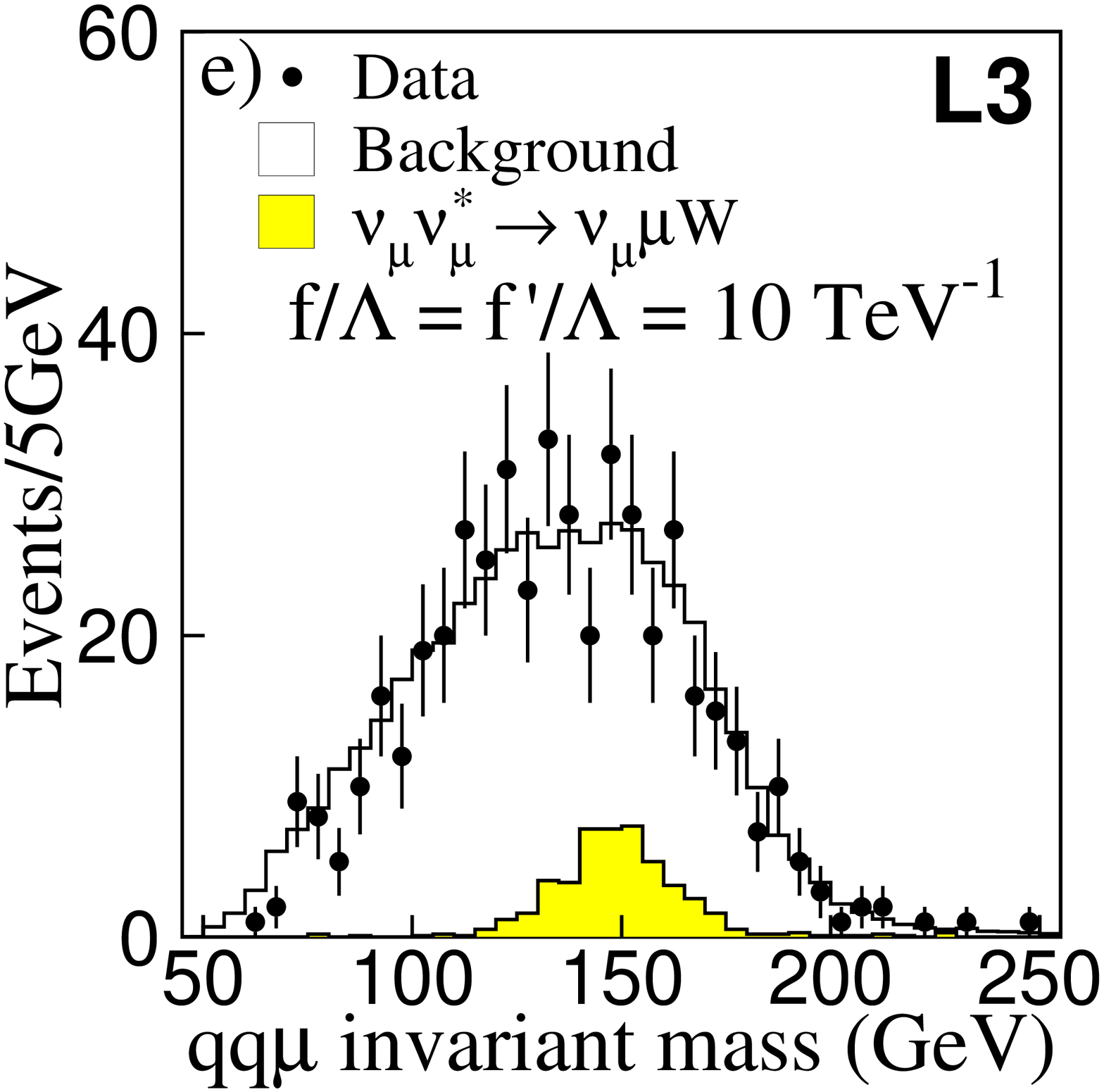}
    \includegraphics[width=0.40\textwidth]{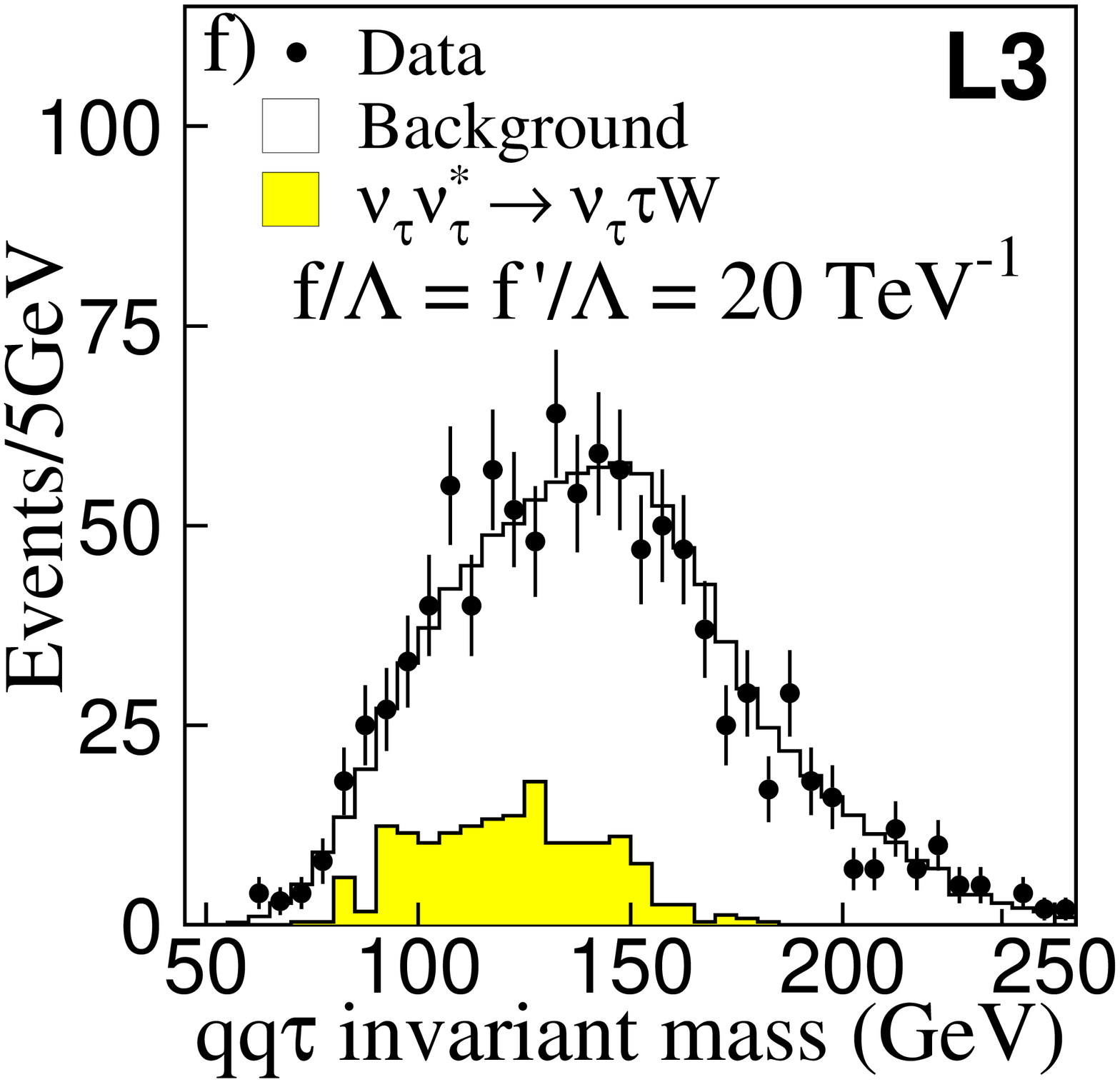}
  \end{center}
  \icaption{Lepton recoil mass distributions for the 
           a) $\rm qqe\nu$, b) $\rm qq\mu\nu$
           and c) $\rm qq\tau\nu$ selections. 
           Invariant mass distributions for d) $\rm qqe\nu$, 
           e) $\rm qq\mu\nu$ and f) $\rm qq\tau\nu$  selected events.
           The expected signal
           for an excited lepton with a mass of $150 \GeV$ 
           is shown together with data and Standard
           Model background.
           The signals are plotted for the arbitrary choice of couplings 
           displayed in the Figures.
  \label{fig:weak}}
\end{figure}

The experimental signatures of
\ee$\rm\longrightarrow \ell \ell^* \longrightarrow \ell \nu  W$ and
\ee$\rm\longrightarrow \nu  \nu^*  \longrightarrow \nu  \ell W$ 
are similar to W pair production. 
For the final states containing a pair of hadronic jets from the 
W decay, the $\rm qq\ell\nu$
selections described above are applied.
However, in order to be sensitive to charged excited leptons with
mass close to the W mass, and to excited neutrinos with mass close
to $\sqrt{s}$, where the final state neutrino carries low momentum,
no cuts on the missing momentum or visible energy are applied.
Figures~\ref{fig:weak}a-c show the recoil mass against the identified 
lepton in the $\rm qqe\nu$, $\rm qq\mu\nu$ and $\rm qq\tau\nu$ selections, 
indicating the mass of the $\ell^*$ candidates.
Figures~\ref{fig:weak}d-f show the event invariant mass for the 
$\rm qqe\nu$, $\rm qq\mu\nu$ and $\rm qq\tau\nu$ selected events,
indicating the mass of the $\nu^*$ candidates. 

For the cases in which the final state lepton escapes undetected, the
$\rm qq\ell\nu$ selections are complemented with a $\rm qq\nu\nu$ 
selection which requires two acoplanar hadronic jets with invariant 
mass between $60 \GeV$ and $100 \GeV$, visible energy 
smaller than $150 \GeV$ and recoil mass below $70 \GeV$.

If the W boson decays leptonically, the event contains two charged
leptons and two neutrinos, $\rm \ell \ell^* \longrightarrow \ell \nu  W 
\longrightarrow \ell \nu  \ell \nu$ or $\rm \nu  \nu^*  \longrightarrow 
\nu  \ell W \longrightarrow \nu  \ell \ell \nu$. 
For this signature, low multiplicity events containing two charged leptons 
with acoplanarity between $10^\circ$ and $170^\circ$ are selected.
The energy of the most energetic lepton must exceed $20 \GeV$ and the second 
lepton must exceed $3 \GeV$.
To reject the background from two-photon interactions, the transverse 
missing momentum must be greater than $10 \GeV$. 

%
%
\subsection{Neutral-Current Decays}

The processes 
\ee$\rm\longrightarrow \ell \ell^* \longrightarrow \ell \ell  Z$ and
\ee$\rm\longrightarrow \nu  \nu^*  \longrightarrow \nu  \nu Z$ 
give rise to four visible experimental signatures according to
the decay channel of the Z boson: $\rm qq\ell\ell$, $\rm qq\nu\nu$,
$\ell\ell\ell\ell$ and $\ell\ell\nu\nu$. 
The $\rm qq\ell\ell$, $\rm qq\nu\nu$ and $\ell\ell\nu\nu$ signatures are 
investigated making use of the same selections used in the search for 
single excited leptons decaying via charged current.
For the $\ell\ell\ell\ell$ final state, low multiplicity events with 
at least three charged leptons are selected. 
The invariant mass of a pair of these leptons 
must be in the range from $80 \GeV$ to $100 \GeV$.

%
%
\section{Results}

The number of events observed in the data for each of the selections described 
above is consistent with the Standard Model background expectation, 
as shown in Table~\ref{tab:selections}. 
For some signatures of excited leptons, a sizable signal efficiency is 
achieved with very low background, giving sensitivity to
production cross sections as low as 0.1 pb. 

For each flavour of excited lepton, all selections are combined to derive 
an upper limit to the signal cross section, taking into account the 
luminosity, the branching fractions and the efficiencies. 
The results from the four centre-of-mass energies analysed are 
presented together, but are treated separately for the calculation 
of limits, as the signal cross sections depend on $\sqrt{s}$.
These combinations include the results from 
$\sqrt{s} = 189 \GeV$\cite{L3_189}.
The limit is set at 95\% confidence level, using Bayesian statistics 
and assuming a flat positive {\it a priori} distribution for the 
signal cross section.
Two different scenarios are considered in order to present limits. 
In the first one, $f=f'$, the radiative decay is allowed for
charged excited leptons whereas it is forbidden for excited neutrinos.
In the second one, $f=-f'$, the situation is the opposite: 
the radiative decay is forbidden for charged excited leptons and 
is allowed for excited neutrinos.

In the case of pair production, lower mass limits are derived from 
the cross section upper bounds.
A scan is performed for all the possible relative values of $f$ and $f'$.
For each value, the corresponding decay fractions 
$\rm \ell^* \ell^* \longrightarrow \ell \ell \gamma \gamma$,
$\rm \ell^* \ell^* \longrightarrow \nu \nu W W$ and
$\rm \ell^* \ell^* \longrightarrow \ell \gamma \nu W$ or
$\nu^* \nu^* \longrightarrow \nu \nu \gamma \gamma$,
$\rm \nu^* \nu^* \longrightarrow \ell \ell W W$ and
$\rm \nu^* \nu^* \longrightarrow \nu \gamma \ell W$ 
in the case of excited neutrinos
are calculated, and a mass limit is set.  
In Table~\ref{tab:limits} the limits corresponding to the $f=f'$
and $f=-f'$ scenarios are presented together with the lowest limit 
obtained in the scan. This limit corresponds to the ratio of the
couplings  $f/f'$ that, given the  
signal efficiencies and the number of expected candidates, maximise the
compatibility of the data with the existence of excited leptons.
The lowest mass limits correspond to the branching ratios that, given 
the signal efficiencies and the number of expected candidates, maximise 
the compatibility of the data with the existence of excited leptons. 

\begin{figure}[htb]
  \begin{center}
    \includegraphics[width=0.49\textwidth]{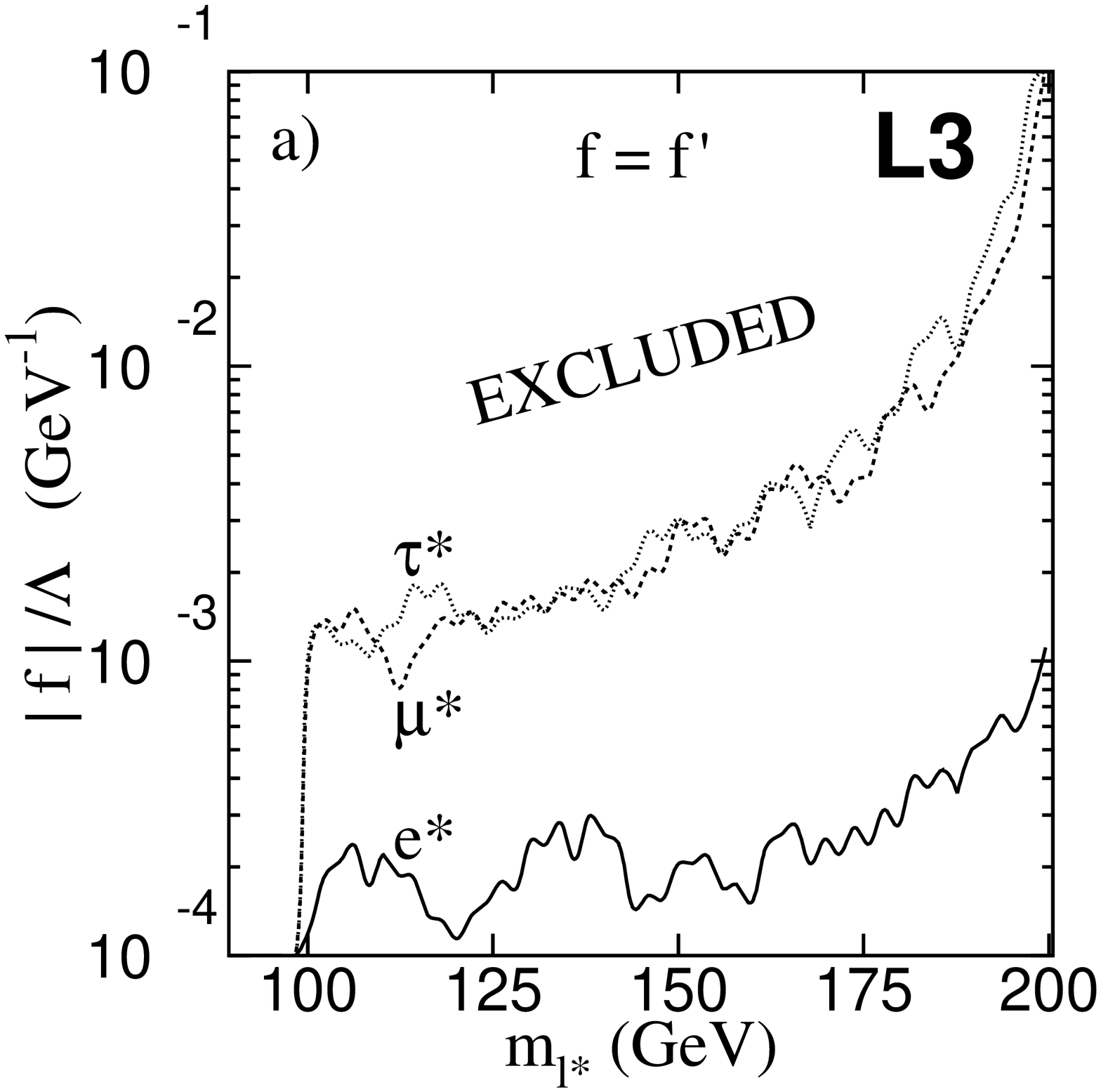}
    \includegraphics[width=0.49\textwidth]{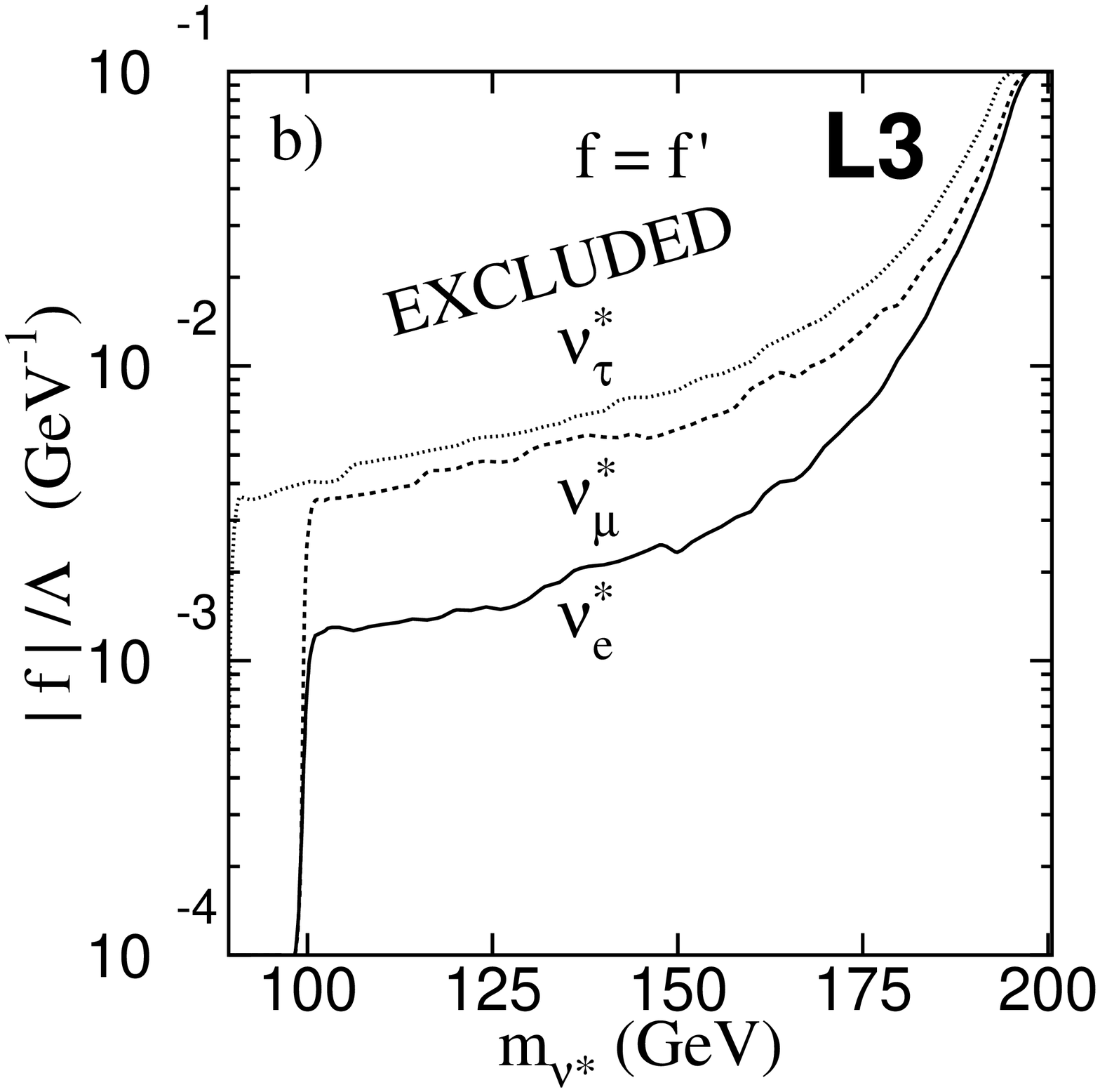}
    \includegraphics[width=0.49\textwidth]{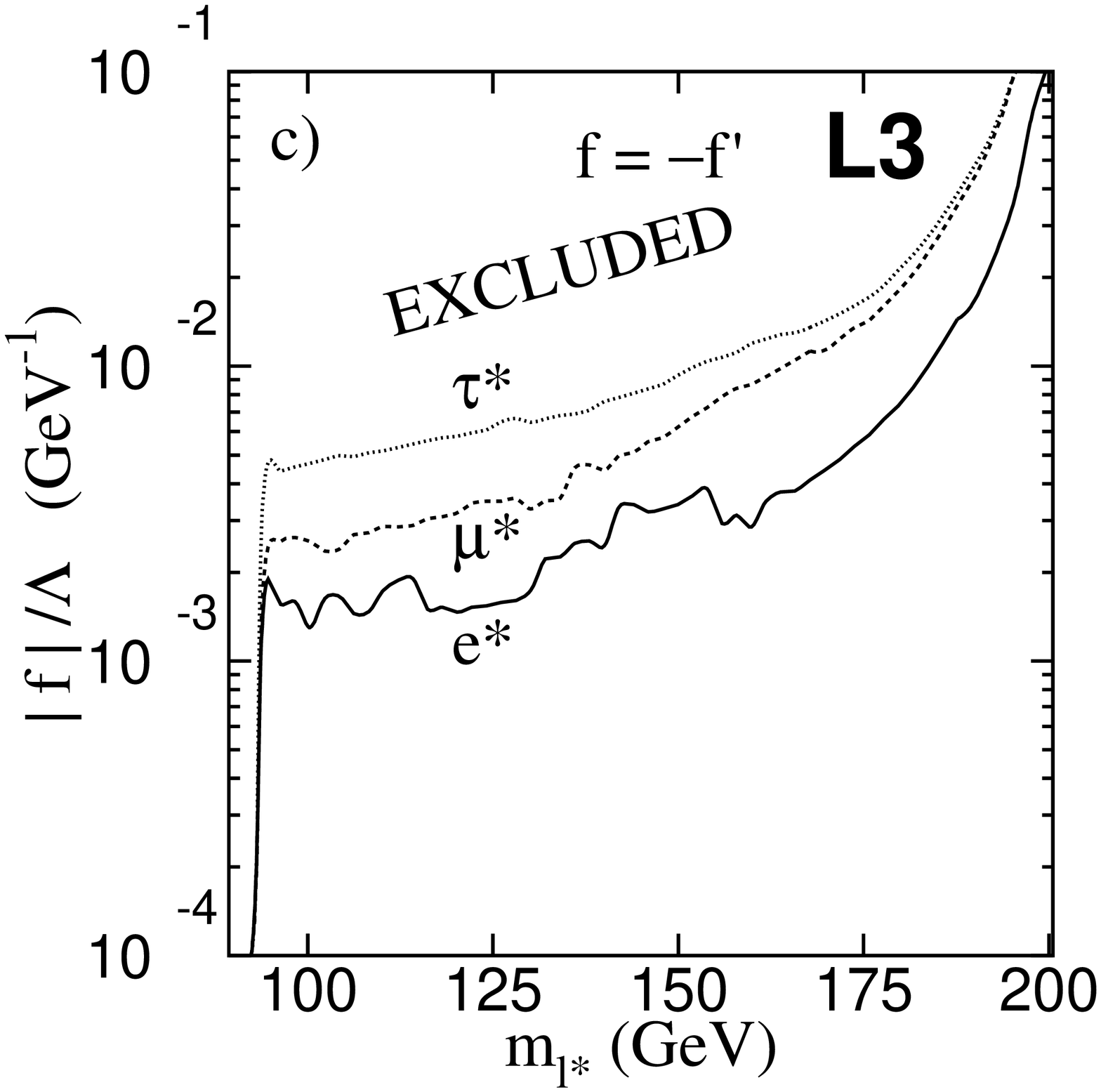}
    \includegraphics[width=0.49\textwidth]{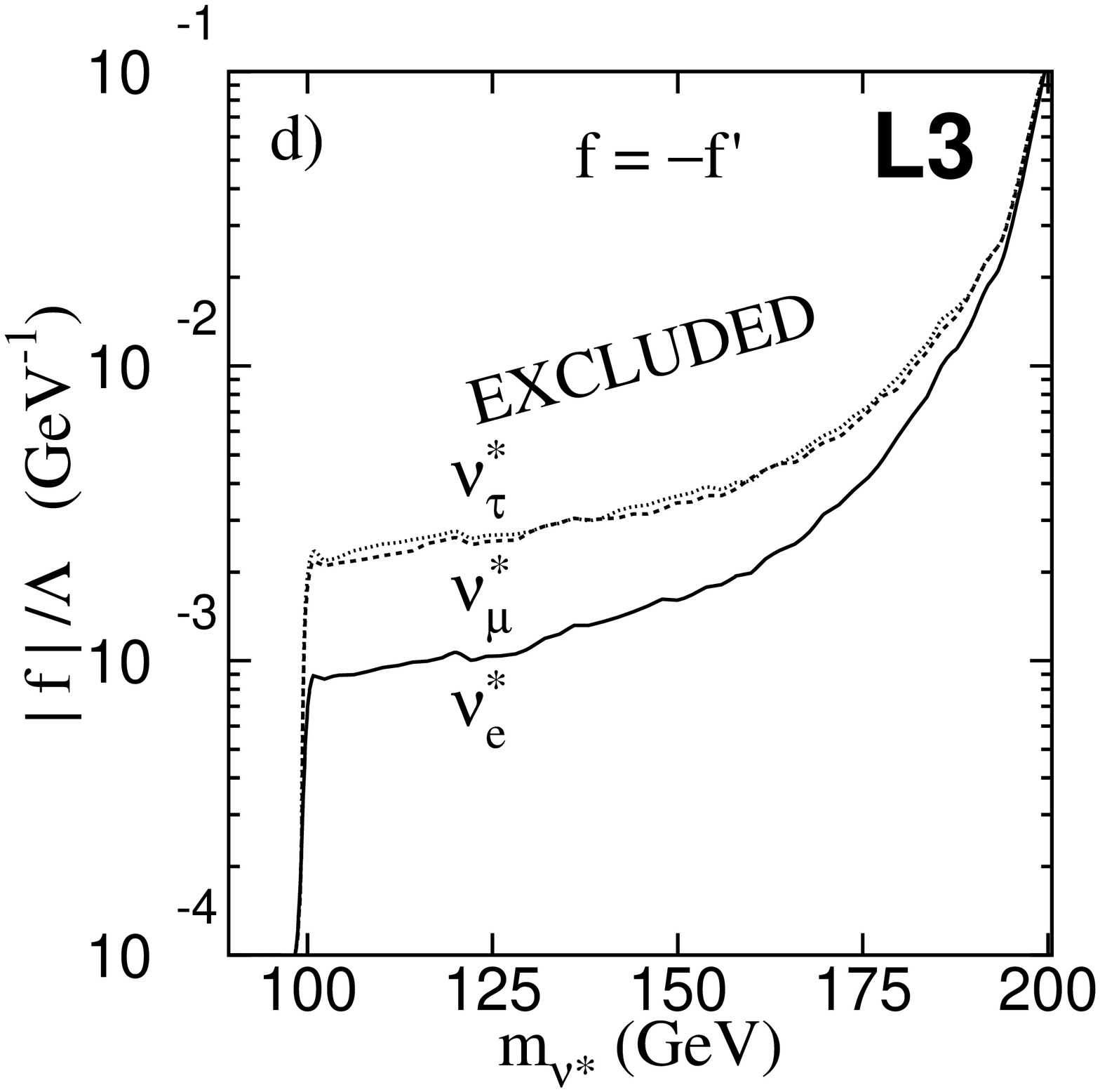}
  \end{center}
  \icaption{95\% confidence level upper limits on the effective coupling 
           ${\textstyle |f|} / {\textstyle \Lambda}$,
           as a function of the excited lepton mass  with $f=f'$:
       a) $\rm e^*$, $\mu^*$ and $\tau^*$,
       b) $\rm \nu_e^*$, $\nu_{\mu}^*$ and $\nu_{\tau}^*$,
           and with $f=-f'$:
       c) $\rm e^*$, $\mu^*$ and $\tau^*$,
       d) $\rm \nu_e^*$, $\nu_{\mu}^*$ and $\nu_{\tau}^*$.
  \label{fig:limites}}
\end{figure}

In the case of single production searches, an upper limit to the cross
section is obtained as a function of the excited lepton mass.
The distributions of the variables presented in Figures~\ref{fig:rad} 
and~\ref{fig:weak} are investigated for each mass hypothesis.
The number of candidates found in data and expected
from background, as well as the efficiency are calculated in the range where
the signal is expected.
For the charged-current decay with the W decaying leptonically, and
for the neutral-current decay, all the selected events are considered as 
candidates for all the mass hypotheses.
The results from all selections are combined to derive a cross section 
limit which is interpreted in terms of an upper limit to the effective 
coupling constant $|f|/\Lambda$.
Figure~\ref{fig:limites} shows the upper limits to the effective
coupling $|f|/\Lambda$ for charged excited leptons and 
excited neutrinos for the $f=f'$ and $f=-f'$  scenarios. 
The left-hand edge of the curves indicates the lower mass limit 
derived from pair production searches whereas the rise of the curves 
in the high excited lepton mass region reflects the lack of experimental
sensitivity due to the low expected signal cross section.
The limits corresponding to charged excited leptons in the $f=f'$ scenario
and excited neutrinos in the $f=-f'$ scenario are derived mainly from the
radiative decay searches, highly sensitive to the signal. These limits
are therefore more stringent than the limits obtained in the complementary
scenarios in which the radiative decays are forbidden.
Moreover, the limits corresponding to excited leptons of the first generation
are significantly lower due to the higher cross section explained by the
$t$-channel contribution.

The systematic uncertainties are conservatively taken into account for the
limit calculations by lowering the background expectation and the signal 
efficiency in all the selections.
A variation close to 2.6\% in the background, depending on the selection,
accounts for the uncertainties related to background cross sections, limited
Monte Carlo statistics, detector simulation and selection efficiency. 
A similar variation in the signal efficiency accounts for the 
uncertainties coming from signal Monte Carlo statistics, the extrapolation
to different centre-of-mass energies and excited lepton masses and 
detector simulation. The effect of systematics is quite significant for
the limits extracted from high statistics selections,
as in the case of $\nu_\tau^*$ in the $f=f'$ scenario.

In conclusion, lower mass limits as high as $98.3 \GeV$ 
are set for any value of the couplings.  Upper limits
on the couplings $|f|/\Lambda$ ranging from $10^{-1}$ to $10^{-4} \GeV^{-1}$ 
are derived in the mass range from $90 \GeV$ to $200 \GeV$.
%
%
\section*{Acknowledgements}

We thank the CERN accelerator divisions for the continuous and 
successful upgrade of the LEP machine and its excellent performance. 
We acknowledge the contributions of the engineers and technicians who
have participated in the construction and maintenance of this experiment.

%
%

%
%
  \newpage
\typeout{   }     
\typeout{Using author list for paper 226 -- ? }
\typeout{$Modified: Tue Sep  5 19:04:46 2000 by smele $}
\typeout{!!!!  This should only be used with document option a4p!!!!}
\typeout{   }
%
%
%
%
%
%

\newcount\tutecount  \tutecount=0
\def\tutenum#1{\global\advance\tutecount by 1 \xdef#1{\the\tutecount}}
\def\tute#1{$^{#1}$}
\tutenum\aachen            
\tutenum\nikhef            
\tutenum\mich              
\tutenum\lapp              
\tutenum\basel             
\tutenum\lsu               
\tutenum\beijing           
\tutenum\berlin            
\tutenum\bologna           
\tutenum\tata              
\tutenum\ne                
\tutenum\bucharest         
\tutenum\budapest          
\tutenum\mit               
\tutenum\debrecen          
\tutenum\florence          
\tutenum\cern              
\tutenum\wl                
\tutenum\geneva            
\tutenum\hefei             
\tutenum\seft              
\tutenum\lausanne          
\tutenum\lecce             
\tutenum\lyon              
\tutenum\madrid            
\tutenum\milan             
\tutenum\moscow            
\tutenum\naples            
\tutenum\cyprus            
\tutenum\nymegen           
\tutenum\caltech           
\tutenum\perugia           
\tutenum\cmu               
\tutenum\prince            
\tutenum\rome              
\tutenum\peters            
\tutenum\potenza           
\tutenum\riverside         
\tutenum\salerno           
\tutenum\ucsd              
\tutenum\santiago          
\tutenum\sofia             
\tutenum\korea             
\tutenum\alabama           
\tutenum\utrecht           
\tutenum\purdue            
\tutenum\psinst            
\tutenum\zeuthen           
\tutenum\eth               
\tutenum\hamburg           
\tutenum\taiwan            
\tutenum\tsinghua          

{
\parskip=0pt
\noindent
{\bf The L3 Collaboration:}
\ifx\selectfont\undefined
 \baselineskip=10.8pt
 \baselineskip\baselinestretch\baselineskip
 \normalbaselineskip\baselineskip
 \ixpt
\else
 \fontsize{9}{10.8pt}\selectfont
\fi
\medskip
\tolerance=10000
\hbadness=5000
\raggedright
\hsize=162truemm\hoffset=0mm
\def\r{\rlap,}
\noindent

M.Acciarri\r\tute\milan\
P.Achard\r\tute\geneva\ 
O.Adriani\r\tute{\florence}\ 
M.Aguilar-Benitez\r\tute\madrid\ 
J.Alcaraz\r\tute\madrid\ 
G.Alemanni\r\tute\lausanne\
J.Allaby\r\tute\cern\
A.Aloisio\r\tute\naples\ 
M.G.Alviggi\r\tute\naples\
G.Ambrosi\r\tute\geneva\
% ...
\newpage
\begin{list}{A}{\itemsep=0pt plus 0pt minus 0pt\parsep=0pt plus 0pt minus 0pt
                \topsep=0pt plus 0pt minus 0pt}
\item[\aachen]
 I. Physikalisches Institut, RWTH, D-52056 Aachen, FRG$^{\S}$\\
 III. Physikalisches Institut, RWTH, D-52056 Aachen, FRG$^{\S}$
\item[\nikhef] National Institute for High Energy Physics, NIKHEF, 
     and University of Amsterdam, NL-1009 DB Amsterdam, The Netherlands
\item[\mich] University of Michigan, Ann Arbor, MI 48109, USA
\item[\lapp] Laboratoire d'Annecy-le-Vieux de Physique des Particules, 
     LAPP,IN2P3-CNRS, BP 110, F-74941 Annecy-le-Vieux CEDEX, France
\item[\basel] Institute of Physics, University of Basel, CH-4056 Basel,
     Switzerland
\item[\lsu] Louisiana State University, Baton Rouge, LA 70803, USA
\item[\beijing] Institute of High Energy Physics, IHEP, 
  100039 Beijing, China$^{\triangle}$ 
\item[\berlin] Humboldt University, D-10099 Berlin, FRG$^{\S}$
\item[\bologna] University of Bologna and INFN-Sezione di Bologna, 
     I-40126 Bologna, Italy
\item[\tata] Tata Institute of Fundamental Research, Bombay 400 005, India
\item[\ne] Northeastern University, Boston, MA 02115, USA
\item[\bucharest] Institute of Atomic Physics and University of Bucharest,
     R-76900 Bucharest, Romania
\item[\budapest] Central Research Institute for Physics of the 
     Hungarian Academy of Sciences, H-1525 Budapest 114, Hungary$^{\ddag}$
\item[\mit] Massachusetts Institute of Technology, Cambridge, MA 02139, USA
\item[\debrecen] KLTE-ATOMKI, H-4010 Debrecen, Hungary$^\P$
\item[\florence] INFN Sezione di Firenze and University of Florence, 
     I-50125 Florence, Italy
\item[\cern] European Laboratory for Particle Physics, CERN, 
     CH-1211 Geneva 23, Switzerland
\item[\wl] World Laboratory, FBLJA  Project, CH-1211 Geneva 23, Switzerland
\item[\geneva] University of Geneva, CH-1211 Geneva 4, Switzerland
\item[\hefei] Chinese University of Science and Technology, USTC,
      Hefei, Anhui 230 029, China$^{\triangle}$
\item[\lausanne] University of Lausanne, CH-1015 Lausanne, Switzerland
\item[\lecce] INFN-Sezione di Lecce and Universit\`a Degli Studi di Lecce,
     I-73100 Lecce, Italy
\item[\lyon] Institut de Physique Nucl\'eaire de Lyon, 
     IN2P3-CNRS,Universit\'e Claude Bernard, 
     F-69622 Villeurbanne, France
\item[\madrid] Centro de Investigaciones Energ{\'e}ticas, 
     Medioambientales y Tecnolog{\'\i}cas, CIEMAT, E-28040 Madrid,
     Spain${\flat}$ 
\item[\milan] INFN-Sezione di Milano, I-20133 Milan, Italy
\item[\moscow] Institute of Theoretical and Experimental Physics, ITEP, 
     Moscow, Russia
\item[\naples] INFN-Sezione di Napoli and University of Naples, 
     I-80125 Naples, Italy
\item[\cyprus] Department of Natural Sciences, University of Cyprus,
     Nicosia, Cyprus
\item[\nymegen] University of Nijmegen and NIKHEF, 
     NL-6525 ED Nijmegen, The Netherlands
\item[\caltech] California Institute of Technology, Pasadena, CA 91125, USA
\item[\perugia] INFN-Sezione di Perugia and Universit\`a Degli 
     Studi di Perugia, I-06100 Perugia, Italy   
\item[\cmu] Carnegie Mellon University, Pittsburgh, PA 15213, USA
\item[\prince] Princeton University, Princeton, NJ 08544, USA
\item[\rome] INFN-Sezione di Roma and University of Rome, ``La Sapienza",
     I-00185 Rome, Italy
\item[\peters] Nuclear Physics Institute, St. Petersburg, Russia
\item[\potenza] INFN-Sezione di Napoli and University of Potenza, 
     I-85100 Potenza, Italy
\item[\riverside] University of Californa, Riverside, CA 92521, USA
\item[\salerno] University and INFN, Salerno, I-84100 Salerno, Italy
\item[\ucsd] University of California, San Diego, CA 92093, USA
\item[\santiago] Dept. de Fisica de Particulas Elementales, Univ. de Santiago,
     E-15706 Santiago de Compostela, Spain
\item[\sofia] Bulgarian Academy of Sciences, Central Lab.~of 
     Mechatronics and Instrumentation, BU-1113 Sofia, Bulgaria
\item[\korea]  Laboratory of High Energy Physics, 
     Kyungpook National University, 702-701 Taegu, Republic of Korea
\item[\alabama] University of Alabama, Tuscaloosa, AL 35486, USA
\item[\utrecht] Utrecht University and NIKHEF, NL-3584 CB Utrecht, 
     The Netherlands
\item[\purdue] Purdue University, West Lafayette, IN 47907, USA
\item[\psinst] Paul Scherrer Institut, PSI, CH-5232 Villigen, Switzerland
\item[\zeuthen] DESY, D-15738 Zeuthen, 
     FRG
\item[\eth] Eidgen\"ossische Technische Hochschule, ETH Z\"urich,
     CH-8093 Z\"urich, Switzerland
\item[\hamburg] University of Hamburg, D-22761 Hamburg, FRG
\item[\taiwan] National Central University, Chung-Li, Taiwan, China
\item[\tsinghua] Department of Physics, National Tsing Hua University,
      Taiwan, China
\item[\S]  Supported by the German Bundesministerium 
        f\"ur Bildung, Wissenschaft, Forschung und Technologie
\item[\ddag] Supported by the Hungarian OTKA fund under contract
numbers T019181, F023259 and T024011.
\item[\P] Also supported by the Hungarian OTKA fund under contract
  numbers T22238 and T026178.
\item[$\flat$] Supported also by the Comisi\'on Interministerial de Ciencia y 
        Tecnolog{\'\i}a.
\item[$\sharp$] Also supported by CONICET and Universidad Nacional de La Plata,
        CC 67, 1900 La Plata, Argentina.
\item[$\diamondsuit$] Also supported by Panjab University, Chandigarh-160014, 
        India.
\item[$\triangle$] Supported by the National Natural Science
  Foundation of China.
\end{list}
}
\vfill


%
%

\begin{table}[th]
\begin{center}
\begin{tabular}{|rcl||c|c||c|c|}\hline
  & &    &  \multicolumn{4}{|c|}{Branching Ratios} \\
\cline{4-7}
\multicolumn{3}{|c||}{Decay}    &  \multicolumn{2}{|c||}{$M=95\GeV$} &
          \multicolumn{2}{|c|}{$M=190\GeV$}    \\
         \cline{4-7}
\multicolumn{3}{|c||}{Channel}  & $f=f'$ & $f=-f'$ & $f=f'$ & $f=-f'$  \\
\hline
$\rm \ell^*$ & $ \longrightarrow $ & $\rm \ell \gamma$ &
          81\%   &    --           &  36\%   &   --       \\
$\rm \ell^* $ & $\longrightarrow $ & $\rm \nu W      $ &
          19\%   &   93\%          &  55\%   &  63\%      \\
$\rm \ell^* $ & $\longrightarrow $ & $\rm \ell Z     $ &
          ~0\%   &   ~7\%          &  ~9\%   &  37\%      \\
\hline
\hline
$\rm \nu^*  $ & $\longrightarrow $ & $\rm \nu  \gamma$ &
           --    &   81\%          &   --    &  36\%      \\
$\rm \nu^*  $ & $\longrightarrow $ & $\rm \ell W     $ &
          93\%   &   19\%          &  63\%   &  55\%      \\
$\rm \nu^*  $ & $\longrightarrow $ & $\rm \nu  Z     $ &
          ~7\%   &   ~0\%          &  37\%   &  ~9\%      \\
\hline
\end{tabular}
\icaption{Predicted branching ratios for charged and neutral
      excited lepton decays, for different choices of masses
              and couplings.
    \label{tab:bratio}}
  \end{center}
\end{table}

\begin{table}[th]
\begin{center}
\begin{tabular}{|c|rcl|c|c|c||rcl|c|c|c|}
\hline

& \multicolumn{6}{|c||}{Charged Excited Leptons} & 
  \multicolumn{6}{c|}{Excited Neutrinos}                       \\

\cline{2-13} 
& \multicolumn{3}{|c|}{Signal} & $N_D$ & $N_B$ & $\epsilon$  
& \multicolumn{3}{|c|}{Signal} & $N_D$ & $N_B$ & $\epsilon$    \\ 

\hline \hline \rule{0pt}{13pt} 
& $\rm e^* e^* $ & $\longrightarrow$ & $\rm e e \gamma \gamma $                
&    1  &    0.8  &   46\%
& $\rm \nu_{e}^* \nu_{e}^* $ & $\longrightarrow$ & $\rm e e W W$
&    0  &    0.2  &   19\%            \\

\rule{0pt}{13pt} 
& $\rm \mu^* \mu^* $ & $\longrightarrow$ & $\rm \mu \mu \gamma \gamma$ 
&    1  &    0.5  &   44\%
& $\rm \nu_{\mu}^* \nu_{\mu}^* $ & $\longrightarrow$ & $\rm \mu \mu W W$
&    0  &    0.7  &   22\%            \\

\rule{0pt}{13pt} 
& $\rm \tau^* \tau^* $ & $\longrightarrow$ & $\rm \tau \tau \gamma \gamma$ 
&    0  &    0.2  &   43\%
& $\rm \nu_{\tau}^* \nu_{\tau}^* $ & $\longrightarrow$ & $\rm \tau \tau W W$
& 2978  & 2993    &   69\%            \\

\cline{2-13} \rule{0pt}{13pt} 
& $\rm \ell^* \ell^* $ & $\longrightarrow$ & $\rm \nu \nu W W$
& 2978  & 2993    &   65\%
& $\rm \nu^*\nu^* $ & $\longrightarrow$ & $\rm \nu \nu \gamma \gamma$
&    2  &    1.9  &   45\%            \\

\cline{2-13} \rule{0pt}{13pt} 
& $\rm e^* e^* $ & $\longrightarrow$ & $\rm e \gamma \nu W $
&   13  &   11    &   57\%
& $\rm \nu_{e}^* \nu_{e}^* $ & $\longrightarrow$ & $\rm \nu \gamma e W$
&       &         &   38\%            \\

\rule{0pt}{13pt} 
& $\rm \mu^* \mu^* $ & $\longrightarrow$ & $\rm \mu \gamma \nu W $
&    5  &    8    &   43\%
& $\rm \nu_{\mu}^* \nu_{\mu}^* $ & $\longrightarrow$ & $\rm \nu \gamma \mu W $
&   18  &   12    &   28\%            \\

\rule{0pt}{13pt} \raisebox{4.0mm}[0mm][0mm]
{\rotatebox{90} {Pair Production}} 
& $\rm \tau^* \tau^* $ & $\longrightarrow$ & $\rm \tau \gamma \nu W $
&   56  &   45    &   37\%
& $\rm \nu_{\tau}^* \nu_{\tau}^* $ & $\longrightarrow$ & $\rm \nu \gamma \tau W $
&       &         &   19\%            \\

\hline \hline \rule{0pt}{13pt} 
& $\rm e e^* $ & $\longrightarrow$ & $\rm e e \gamma$        
&  682  &  714    &   61\%
&                                           
&       &         &     &&            \\

\rule{0pt}{13pt} 
& $\rm \mu \mu^* $ & $\longrightarrow$ & $\rm \mu \mu \gamma$   
&   72  &   67    &   61\%
& $\rm \nu \nu^* $ & $\longrightarrow$ & $\rm \nu \nu \gamma$   
&  230  &  232    &   64\%            \\

\rule{0pt}{13pt} 
& $\rm \tau \tau^* $ & $\longrightarrow$ & $\rm \tau \tau \gamma$   
&   50  &   69    &   42\%
&                                                  
&       &         &      &&           \\

\cline{2-13} \rule{0pt}{13pt} 
& $\rm e e^* $ & $\longrightarrow$ & $\rm e \nu_e W$                
&  728  &  764    &   67\%
& $\rm \nu_{e} \nu_{e}^* $ & $\longrightarrow$ & $\rm \nu_{e} e W$  
&  728  &  764    &   65\%            \\

\rule{0pt}{13pt} 
& $\rm \mu \mu^* $ & $\longrightarrow$ & $\rm \mu \nu_{\mu} W$             
&  684  &  694    &   60\%
& $\rm \nu_{\mu} \nu_{\mu}^* $ & $\longrightarrow$ & $\rm \nu_{\mu} \mu W$ 
&  684  &  694    &   65\%            \\

\rule{0pt}{13pt} 
& $\rm \tau \tau^* $ & $\longrightarrow$ & $\rm \tau \nu_{\tau} W$
& 1365  & 1377    &   55\%
& $\rm \nu_{\tau} \nu_{\tau}^* $ & $\longrightarrow$ & $\rm \nu_{\tau} \tau W$
& 1365  & 1377    &   56\%            \\

\cline{2-13} \rule{0pt}{13pt} 
& $\rm e e^* $ & $\longrightarrow$ & $\rm e e Z$                    
&  663  &  692    &   58\%
&                                                  
&       &         &       &&          \\

\rule{0pt}{13pt} 
& $\rm \mu \mu^* $ & $\longrightarrow$ & $\rm \mu \mu Z$            
&  608  &  617    &   52\%
& $\rm \nu \nu^* $ & $\longrightarrow$ & $\rm \nu \nu Z$            
&  350  &  357    &   29\%            \\

\rule{0pt}{13pt} \raisebox{9.0mm}[0mm][0mm]
{\rotatebox{90} {Single Production}} 
& $\rm \tau \tau^* $ & $\longrightarrow$ & $\rm \tau \tau Z$        
& 1307  & 1308    &   34\%
&                                                  
&       &         &        &&         \\

\hline
\end{tabular}
\icaption{Number of candidates $N_D$, number of background events $N_B$, 
          and average signal efficiencies $\epsilon$, in the pair
          production (upper part) and the single production
          (lower part) searches.
\label{tab:selections}}
\end{center}
\end{table}
\newpage

\begin{table}[th]
\begin{center}
\begin{tabular}{|c||c|c|c|}
\hline
    Excited           &  \multicolumn{3}{|c|}{95\% CL Mass Limit (\GeV)} \\
    \cline{2-4}
    Lepton              & $f=f'$ & $f=-f'$& Any Coupling \\
    \hline
    $\rm \e^*         $ & 100.0  &  93.4  & 93.3 \\
    $\rm \mu^*        $ & 100.2  &  93.4  & 93.4 \\
    $\rm \tau^*       $ & ~99.8  &  93.4  & 92.2 \\
    \hline
    $\rm \nu^*_e      $ & ~99.1  &  99.4  & 98.2 \\
    $\rm \nu^*_{\mu}  $ & ~99.3  &  99.4  & 98.3 \\
    $\rm \nu^*_{\tau} $ & ~90.5  &  99.4  & 87.8 \\
    \hline
\end{tabular}
\icaption{95\% confidence level lower mass limits for the different
          excited leptons obtained from pair production searches. For each
          flavour, the mass limits for $f=f'$, $f=-f'$ and for the
          coupling independent case, are shown.
\label{tab:limits}}
\end{center}
\end{table}

\clearpage
%
%

\clearpage

\clearpage

\clearpage

\end{document}